\documentclass[12pt,letterpaper]{article}
\pdfoutput=1
\usepackage{graphicx, epsfig, color,cite}
\usepackage{amsmath}
\usepackage{amssymb}
\usepackage{float}
\usepackage{hyperref}
\usepackage{subcaption}
\def\to{\rightarrow}

\def\bi{\begin{itemize}}
\def\ei{\end{itemize}}

\def\tchi{\tilde\chi}

\def\ta{\tilde a}

\def\tst{\tilde t}

\def\tg{\tilde g}

\def\alt{\lesssim}
\def\agt{\gtrsim}
\def\be{\begin{equation}}  
\def\ee{\end{equation}}  
\def\bea{\begin{eqnarray}}  
\def\eea{\end{eqnarray}}

\begin{document}
\begin{titlepage}
\begin{flushright}
OU-HEP-260901
\end{flushright}

\vspace{0.5cm}
\begin{center}
  {\Large \bf Cosmological moduli problem\\
    ameliorated by decaying WIMPs
}
\vspace{1.2cm} \renewcommand{\thefootnote}{\fnsymbol{footnote}}

{\large Howard Baer$^{1}$\footnote[1]{Email: baer@ou.edu },
Vernon Barger$^2$\footnote[2]{Email: barger@pheno.wisc.edu},
Robert Wiley Deal$^2$\footnote[3]{Email: wileydeal@wisc.edu}
}\\ 
\vspace{1.2cm} \renewcommand{\thefootnote}{\arabic{footnote}}
{\it 
$^1$Homer L. Dodge Department of Physics and Astronomy,
University of Oklahoma, Norman, OK 73019, USA \\[3pt]
}
{\it 
$^2$Department of Physics,
University of Wisconsin, Madison, WI 53706, USA \\[3pt]
}
%
\end{center}

\vspace{0.5cm}
\begin{abstract}
\noindent
We investigate the cosmological moduli problem (CMP) in the context of
supersymmetric models where discrete $R$-symmetries are used to
suppress the $\mu$ term, and where $\mu$
is regenerated via Kim--Nilles superpotential operators.
In such theories, a global $U(1)_{PQ}$ emerges as an accidental, approximate
symmetry which solves the strong CP problem.
$R$-parity is also generated as accidental but approximate, where RPV
operators gain a $(f_a/m_P)^n$ suppression.
When $n=1$, the thermally-produced LSPs may decay before BBN in the
early universe, leaving axion-only dark matter of the SUSY DFSZ type.
Meanwhile, the dominant cosmological constraint on light stringy moduli
previously came from the modulus-induced WIMP overproduction problem.
WIMP overproduction is now ameliorated by the RPV WIMP decays,
but the moduli-induced BBN bound may be exacerbated by convolving
long-lived modulus decay with long-lived RPV WIMP decay:
long-lived particle (LLP) cascade decays.
In this context, moduli as light as $m_\phi\sim 130$ TeV may be allowed
which can reconcile naturalness with the presence of stringy moduli
in the early universe.
\end{abstract}
\end{titlepage}

\section{Introduction}
\label{sec:intro}

Kaluza--Klein theory\cite{Bailin:1987jd} is remarkable in that it provides a
putative unification of gravity and electromagnetism (EM) in five spacetime
dimensions, where the
$5-d$ metric tensor contains the $4-d$ metric $g_{\mu\nu}$ plus the EM gauge
potential $A_\mu$ plus a scalar field $\phi$.
The theory assumes the 5th dimension is compactified on a circle leading to
a geometrical origin for the local $U(1)_{EM}$ gauge symmetry of E\&M.
The $\phi$ field is known as the dilaton/radion/graviscalar, and describes the
geometry of the compactification;
as such, it is an example of the emergence of a {\it modulus} field $\phi$
under compactification of extra dimensions.

In string/M-theory formulated in 10 or 11 spacetime
dimensions, once the extra six dimensions\footnote{The extra dimension in $11-d$ M-theory may be compactified on a line segment\cite{Horava:1996ma}
  as in Horava--Witten theory.} are compactified on a Calabi--Yau
space\cite{Candelas:1985en}, a dilaton, along with potentially hundreds
of other moduli
(gravitationally coupled scalar fields describing the size and shape of the
compactified space) are expected to emerge\cite{Conlon:2015uua} in the 4 dimensional effective theory.
A central issue of modern string theory is stabilization of the various moduli.
In type IIB theory, this can be done by a combination\cite{Kachru:2003aw}
of fluxes (for complex structure moduli $U^\alpha$) and non-perturbative effects
(for K\"ahler moduli $T^\beta$). The $U^\alpha$s should gain KK/Planck-scale
masses whilst the $T^\beta$ masses can be much lighter, possibly as light as the
weak scale\cite{Acharya:2010af}.
Light moduli can provide alterations to the standard picture of Big-Bang cosmology, tentatively leading to problematic outcomes in the early universe -- a situation known
as the cosmological moduli problem (CMP)\cite{Coughlan:1983ci,Banks:1993en,deCarlos:1993wie} (for a review, see {\it e.g.} \cite{Kane:2015jia}).

A light modulus field $\phi$, assumed here to be homogeneous and only gravitationally-coupled, obeys the equation of motion
\be
\ddot{\phi}+3H\dot{\phi}+\frac{dV}{d\phi}\simeq 0
\ee
in an expanding universe, where $H\equiv \dot{R}/R$ is the Hubble parameter,
$R$ is the cosmic scale factor and $V(\phi )$ is the $\phi$ field potential
energy which at first approximation\footnote{Although the approximations of
  homogeneity, minimal coupling, and a potential dominated by its mass term are typically good approximations for moduli, the appearance of additional operators (e.g. non-polynomial self-couplings and couplings to kinetic terms) may induce instabilities that trigger rapid non-perturbative decay, entirely analogous to inflationary preheating. These effects are beyond the scope of this work, however see e.g. Refs. \cite{Antusch:2017flz,Price:2026kjs} for work concerning moduli self-couplings and Refs. \cite{Leedom:2024qgr,WileyDeal:2025wgh} and \cite{Giblin:2017wlo} for analysis of moduli couplings with axions and gauge fields, respectively.}
 may be taken as the simple-harmonic
oscillator potential $V\sim\frac{1}{2}m^2 \phi^2$.
At very early times, the $\phi$ field is frozen in time-variation due
to the Hubble friction term but as the universe expands the $\phi$ field begins
to oscillate as cold matter, and then decay.
Since $\phi$ is only gravitationally coupled, it is expected to have a
late-time decay, and if it decays after the onset of Big-Bang nucleosynthesis
(BBN), it threatens the successful predictions of nuclear abundances as
expected from the standard Big Bang cosmology (for a review,
see {\it e.g.} B. Fields {\it et al.} in Ref. \cite{ParticleData:2026uvj}).
If the $\phi$ field is heavy enough (of order $m_\phi \agt 100$ TeV\cite{Kawasaki:1995cy}),
then it may decay before the onset of BBN, thus solving the simplest version
of the CMP.

\subsection{CMP: tension with naturalness}

In supergravity (SUGRA) models, expected to be the low-energy effective field
theory (LE-EFT) from compactified strings,
it is commonly expected that the lightest
modulus gains its mass from SUSY breaking, where 
\be
m_\phi\sim m_{soft}\sim m_{weak}\sim m_{3/2}\sim m_{scalars}\sim m_{hidden}^2/m_P\ \ \ (SUGRA)
\ee
where $m_P$ is the reduced Planck scale and $m_{hidden}\sim 10^{11}$ GeV is a
mass scale associated with hidden sector SUSY breaking\cite{Nilles:2001ap}.
In such a case, one might expect not only the lightest modulus field
$\phi$ to have mass $\agt 100$ TeV, but also all the scalars of the
visible sector.
In the case of the top-squarks, the decoupling solution to the CMP
would induce a serious naturalness problem.
This is true in the case of the most conservative/model-independent
naturalness measure $\Delta_{EW}$\cite{Baer:2012up,Baer:2012cf}
which is based on the MSSM scalar potential
minimization conditions which relate the magnitude of the weak scale to the
magnitude of its SUSY contributions:
\be
m_Z^2/2 = \frac{m_{H_d}^2+\Sigma_d^d -(m_{H_u}^2+\Sigma_u^u )\tan^2\beta}{\tan^2\beta -1}-\mu^2\simeq -m_{H_u}^2-\mu^2-\Sigma_u^u (\tst_{1,2}) .
\label{eq:mzs}
\ee
Here, $m_{H_u}^2$ and $m_{H_d}^2$ are weak scale soft breaking Higgs masses,
$\mu$ is the SUSY conserving Higgs/higgsino mass scale and the
$\Sigma_{u,d}^{u,d}$ terms contain an assortment of loop corrections\cite{Baer:2012cf}, the
most important of which usually come from the top-squarks.
$\Delta_{EW}$ compares the
largest magnitude of the right-hand-side contributions in Eq. \ref{eq:mzs} to $m_Z^2/2$.
Models with $\Delta_{EW}\alt 30$ are considered as natural.
If the weak scale contributions to $m_Z^2/2$ are in the $\sim 100$ TeV range,
as suggested by the decoupling solution to the CMP, then there is a
conflict between naturalness and the requirements of the CMP.
This apparent conflict may be resolved in models of mixed
moduli/anomaly-mediation which occurs in KKLT\cite{Choi:2004sx} models with
flux compactifications where a hierarchy of scales
\be
m_{\phi}\sim 4\pi^2 m_{3/2}\sim (4\pi^2)^2 m_{soft}\ \ \ (KKLT)
\label{eq:kklt}
\ee
can occur\cite{Choi:2005ge}.

Even with a mass hierarchy as given in Eq. \ref{eq:kklt},
a new problem may ensue: the cosmological gravitino problem (CGP)\cite{Kawasaki:2006gs,Nakamura:2006uc,Endo:2006zj,Shuhmaher:2007pv,Dutta:2009uf,Akita:2016usy}.
Normally, gravitinos $\psi_\mu$ are thermally produced at a rate proportional to
$T_R$ in the early universe\cite{Pradler:2006qh} and if heavy enough ($m_{3/2}\agt 5$ TeV\cite{Kawasaki:1994af,Moroi:1995fs}), their decays are BBN safe\cite{Kohri:2005wn,Kawasaki:2008qe}.
However, they can also be produced at large rates by inflaton decay\cite{Kawasaki:2006hm}
or by modulus decay if $m_\phi >2m_{3/2}$\cite{Nakamura:2006uc,Endo:2006zj,Shuhmaher:2007pv}.
In such cases, delayed modulus decays accompanied by delayed gravitino
decays may disrupt BBN.
Additionally, if gravitinos cascade decay to lightest SUSY particles (LSPs)
in $R$-parity conserving (RPC) models, then they may overproduce dark matter.

A third aspect of the CMP arises from direct, non-thermal production of moduli particles via
coherent oscillations, followed by $\phi \to SUSY$ particles,
with the SUSY particles cascading down to LSP dark matter.
In this case, one has the modulus-induced {\it dark matter overproduction
  problem}\cite{Blinov:2014nla}.
In this latter case, including all modulus decays to MSSM particles,
a lightest modulus mass of $m_\phi \agt 5-10$ PeV
seems to be required\cite{Baer:2021zbj,Bae:2022okh,Baer:2023bbn,WileyDeal:2023sry} (1 PeV $=10^3$ TeV).
To avoid the moduli-induced DM overproduction problem without requiring such large masses, an anthropic solution
selecting for anomalously small modulus field strengths $\phi_0\alt 10^{-7}m_P$
has also been suggested\cite{Baer:2021zbj}.

\subsection{Some previous results}
\label{ssec:previous}

In previous papers\cite{Baer:2021zbj,Bae:2022okh}, we have presented 2-body
decay widths for all modulus $\phi$ decays to MSSM particles, including all phase space and mixing effects.
The cases of helicity suppressed and
non-suppressed moduli decay to gravitinos and to gauginos was also presented.
The decays take place via Planck suppressed operators,
and are an extension of earlier work begun by Moroi and
Randall\cite{Moroi:1999zb} who suggested moduli decays to wino-like LSPs
in anomaly-mediated SUSY breaking (AMSB) models\cite{Randall:1998uk,Giudice:1998xp} where there is otherwise
a thermal underabundance of wino-like WIMP dark matter.
It was found that the combined three aspects of the CMP,
\begin{enumerate}
  \item upset of BBN, 
  \item gravitino overproduction, and 
  \item WIMP dark matter overproduction,
\end{enumerate}
lead to very severe constraints on the lightest modulus field mass requiring
$m_{\phi}\agt 5-10$ PeV, which when combined with SUGRA model building,
can lead to a severe conflict with weak scale naturalness, even with a
scale ordering such as in Eq. \ref{eq:kklt}.
A number of solutions were listed, although none seemed overwhelmingly
compelling.

In Ref. \cite{Baer:2022fou}, we extended the MSSM to include axion-like
particles (ALPs) as expected in stringy models such as
large volume compactifications (LVS\cite{Balasubramanian:2005zx}).
If such light ALPs exist, then there may be a fourth aspect to the CMP:
moduli-induced overproduction of dark radiation (DR), as quantified by
the contribution of DR to the effective number of extra neutrinos
in the universe $\Delta N_{\text{eff}}$. At present, this scenario seems disfavored
since the measured value of $N_{\text{eff}}=2.99\pm 0.17$\cite{Planck:2018vyg}
whilst the Standard Model (SM) prediction is for $N_{\text{eff}}=3.044$.
While light ALPs may be expected in LVS, they do not occur in
KKLT-like compactifications\footnote{In KKLT compactifications, modulus stabilization from the non-perturbative superpotential of e.g. gaugino condensation leaves supersymmetry unbroken prior to uplifting to a deSitter minima, resulting in an ALP that possesses a mass comparable to its associated modulus.}.

In Ref. \cite{Baer:2023bbn}, we extended our treatment to include the
SUSY DFSZ axion model (dubbed the PQMSSM) which includes the axion solution
to the strong CP problem and the Kim--Nilles\cite{Kim:1983dt} solution
to the SUSY $\mu$ problem\cite{Bae:2019dgg}.
We evaluated mixed axion/neutralino dark matter production in natural SUSY
via a nine coupled Boltzmann equation treatment that tracked
\begin{enumerate}
  \item neutralinos,
  \item coherent-oscillation (CO)-produced saxions,
  \item thermally produced (TP) and decay-produced (DP) saxions,
  \item TP axinos,
  \item TP and DP axions,
  \item CO-produced axions,
  \item gravitinos,
  \item radiation and
  \item CO-produced lightest modulus field $\phi$.
    \end{enumerate}
Putting the pieces together, it was found that typically
$m_\phi\agt 5-10$ PeV was needed to solve the CMP.
All in all, there is rather severe tension between solving the CMP and
allowing for weak scale naturalness in SUSY models.

\section{Some new theories of SUSY dark matter}

\subsection{All axion dark matter from SUSY with $R$-parity violation}

\subsubsection{Matter parity and $R$-parity}

Traditionally, the view of SUSY dark matter\cite{Jungman:1995df} ignores
problems like the SUSY $\mu$ problem and the strong $CP$ problem,
and introduces $R$-parity conservation on a rather ad hoc basis.
However, $R$-parity is equivalent to matter parity, wherein all
particles carry a multiplicatively conserved charge $P_M=(-1)^{3(B-L)}$.
Under matter parity, matter fields carry $P_M=-1$ whilst Higgs
and gauge fields carry $P_M=+1$.
Thus, under matter parity, only couplings of the form {\it matter-matter-Higgs} are allowed and $P_M$ or $R$-parity violating couplings are forbidden.
The matter parity may be a remnant of a residual $SO(10)$ symmetry wherein
matter fields occupy the 16-dim spinor representation.
$SO(10)$ then only allows $\mathbf{16}$--$\mathbf{16}$--Higgs couplings, the same as matter/$R$-parity.

\subsubsection{Discrete gauge symmetries} 
\label{sssec:dgs}

An attractive alternative is to impose a discrete gauge symmetry, since
gauge symmetries, unlike global symmetries, can be consistent with
quantum gravity.
Discrete gauge symmetries can arise from continuous gauge symmetries, but where
a charge $nq$ object receives a vev which leaves a charge $q$ object
remaining in the low energy effective field theory (LE-EFT).
The result is a discrete $\mathbb{Z}_n$ symmetry\cite{Krauss:1988zc}.
Ib\'{a}\~{n}ez \& Ross\cite{Ibanez:1991pr} examined anomaly cancellation in
theories with $\mathbb{Z}_2$ and $\mathbb{Z}_3$ discrete gauge symmetries
and found baryon triality $B_3$ as an $R$-parity alternative that also allows
$L$ violation.
This was extended by Dreiner {\it et al.}\cite{Dreiner:2005rd} to higher $n$,
where they found proton-hexality $P_6$,
a $\mathbb{Z}_6$ discrete gauge symmetry alternative.

\subsubsection{Discrete $R$-symmetries}

Discrete $R$-symmetries provide a further alternative to $R$-parity.
In Refs. \cite{Lee:2011dya,Chen:2012tia}, all anomaly-free discrete $R$-symmetries
which could suppress the SUSY $\mu$ parameter and $p$-decay operators but
maintain grand unification conditions were tabulated.
For higher $n\sim 12,\ 24$, these $\mathbb{Z}_n^R$ symmetries can be used
to solve the axion quality problem\cite{Baer:2018avn,Bhattiprolu:2021rrj}.
Discrete $R$-symmetries can arise from reduction of 10-d Lorentz symmetry
to 4-d under string compactifications\cite{Nilles:2013lda,Nilles:2017heg}.

\subsubsection{Discrete $R$-symmetries and all axion dark matter}

In Ref. \cite{Baer:2025oid,Baer:2025srs}, discrete $R$ symmetries were used
in a Kim--Nilles type solution to the SUSY $\mu$ problem.
The usual $W\ni \mu H_uH_d$ term is forbidden by some $\mathbb{Z}_n^R$
discrete $R$-symmetry, although non-renormalizable couplings to
PQ-charged fields $X$ and $Y$ are allowed: $W\ni X^pY^qH_uH_d/m_P^{p+q-1}$.
The augmented MSSM then contains an accidental, approximate
$U(1)_{PQ}$ symmetry\cite{Baer:2026jeb} which can be used to solve the
strong $CP$ problem.
The axion, saxion, and axino fields arise as combinations of the
$X$ and $Y$ fields.
The $\mathbb{Z}_n^R$ symmetry also forbids RPV and dim-5 proton-decay
operators.
SUSY breaking at an intermediate hidden sector scale $m_{hidden}$ leads to
vevs $v_X,v_Y\sim 10^{11}$ GeV for the $X$ and $Y$ fields,
thus breaking the discrete $R$-symmetry as well as the accidental PQ symmetry,
and leading to a weak-scale $\mu\sim f_a^2/m_P$ term where the PQ scale
$f_a\sim m_{hidden}$.
It also leads to RPV terms of the
form $W\ni (f_a/m_P)^n QQQ$ (where the $Q$s are the various matter superfield
multiplets). Depending on which $\mathbb{Z}_n^R$ is used, and which PQ charge
assignments are given to the $X$ and $Y$ field, then the induced RPV
couplings may have typically $n=1$ or $n=3$. In the $n=3$ case,
the RPV couplings lead to a SUSY LSP with lifetime longer than the age
of the universe -- essentially stable.
But, for $n=1$ all LSPs can decay before the
onset of BBN, leaving {\it all axion dark matter} in the early universe.
In this case, some additional suppression of RPV operators is needed
(such as $B_3$ or $P_6$ from Subsubsec. \ref{sssec:dgs})
to fully suppress $p$-decay (as is the case in most RPV theories).

In this paper, we examine the impact of the all-axion DM theory on the CMP.
This is a well-motivated scenario which can successfully evade the
strong constraints from the LZ experiment's\cite{LZ:2024zvo}
search for WIMP dark matter.
It also ends up greatly ameliorating the moduli-induced dark matter problem
in that all the WIMPs decay away before the onset of BBN.

\subsection{Natural SUSY with axino as LSP}

A different SUSY dark matter model with decaying lightest neutralinos comes from
the case where the axino $\ta$ is the LSP.
Recently, this scenario has been examined for natural SUSY models with the
MSSM extended by the DFSZ PQ sector.
In this case, the lightest neutralino suffers a delayed decay to mainly
the modes $\tchi_1^0\to \ta+(h\ or\ Z)$, where the decay width
$\Gamma\sim m_{\tchi}^3/f_a^2$.
In this scenario, one expects a mixture of cold axion plus warm/cool
axino dark matter.
In Ref. \cite{Baer:2026wre}, the right amount of relic DM was found with
mainly axino DM at lower $f_a\sim 10^{11}$ GeV and
mainly axion DM with a small population of axinos for higher $f_a\sim 10^{12}$ GeV.

\section{Impact of decaying lightest neutralino on the CMP}

\subsection{Some benchmark parameters}

For our calculations, we adopt two cases.
{\bf Case 1} is inspired by ordinary gravity-mediation, and for this we take
\be
m_{\phi}=2m_{3/2}\sim 20 m_{soft}\ \ \ \ (Case\ 1)
\label{eq:case1}
\ee
where we expect $m_{soft}$ in the tens-of-TeV range (although the benchmark
spectra given below has sparticles spread over two orders of magnitude).
For Case 1, the modulus field decay to gravitinos is effectively closed due to vanishing phase space.

{\bf Case 2} is inspired by KKLT\cite{Kachru:2003aw} flux compactifications\cite{Choi:2005ge},
where instead a scale hierarchy
\be
m_{\phi}\sim (4\pi^2) m_{3/2}\sim (4\pi^2)^2 m_{soft} \ \ \ \ (Case\ 2)
\label{eq:case2}
\ee
is expected.
For Case 2, the lightest modulus $\phi$ can decay to
gravitino pairs, so one may face both the moduli-induced DM overproduction
problem {\it and} the moduli-induced gravitino production problems.  

To illustrate our results, we adopt the same natural SUSY benchmark (BM)
point used in previous works\cite{Bae:2022okh,Baer:2023bbn}.
The BM point comes from the three-extra-parameter non-universal Higgs model
(NUHM3)\cite{Ellis:2002wv,Baer:2005bu} with parameter space
\begin{equation}
m_0(1,2),\ m_0(3),\ m_{1/2},\ A_0,\ \tan\beta ,\ \mu,\ m_A\ \ \ (NUHM3).
\end{equation}
We adopt parameters $m_0(1,2)=10$ TeV for first/second generation matter
scalars, with $m_0(3)=5$ TeV for the third generation.
Also, we take $m_{1/2}=1.2$ TeV, $A_0=-8$ TeV and  $\tan\beta =10$ with $\mu =200$ GeV
and $m_A=2$ TeV.
We generate the spectra using Isajet 7.91\cite{Paige:2003mg}.
With a gluino mass at $m_{\tg}\sim 2.9$ TeV, light top-squark
$m_{\tst_1}\simeq 1.25$ TeV and $m_h=125.3$ GeV, the BM point is in accord
with LHC measurements and sparticle search limits\cite{ATLAS:2024lda,Sekmen:2025bxv}.
The point has electroweak naturalness measure $\Delta_{EW}=20$ so it is
EW natural with no Little Hierarchy Problem (LHP).
The thermally-produced (TP) neutralino abundance from IsaReD\cite{Baer:2002fv}
is $\Omega_{\chi}^{TP}h^2\sim 0.011$ so the putative TP
higgsino-LSP dark matter would be underproduced.

We augment the MSSM with the SUSY DFSZ axion model\cite{Bae:2013hma}
(so strong CP and $\mu$ problems are solved)
and so we take also that $m_s=m_{\ta}=30$ TeV but we leave $f_a$ to vary
along with the axion field misalignment angle $\theta_i$.
Additionally, throughout this work we take the self-coupling of the PQ sector $\xi=0$. 
As was shown in \cite{Baer:2023bbn}, the dark radiation production by saxion decays to axion pairs allowed in the $\xi=1$ limit produces sizeable dark radiation, which we find to be in excess of the currently allowed $\Delta N_{\text{eff}} \lesssim 0.26$ bounds.

\subsection{Energy densities for Case 1}

\begin{figure}[htb!]
\centering
\includegraphics[height=0.4\textheight]{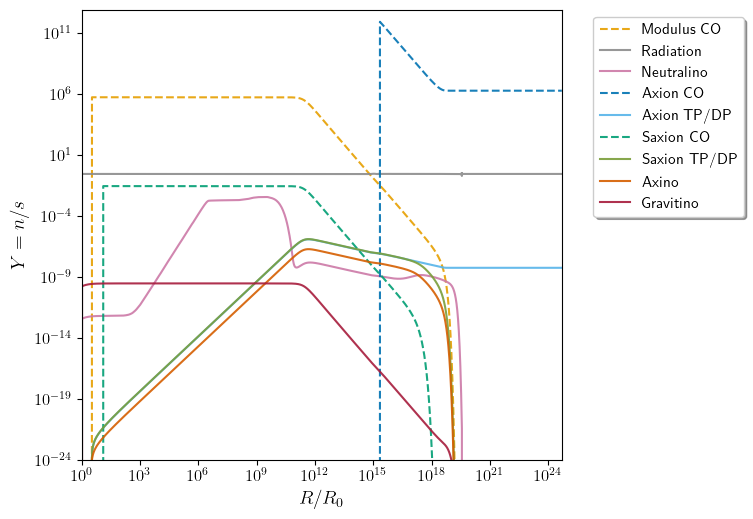}
\includegraphics[height=0.4\textheight]{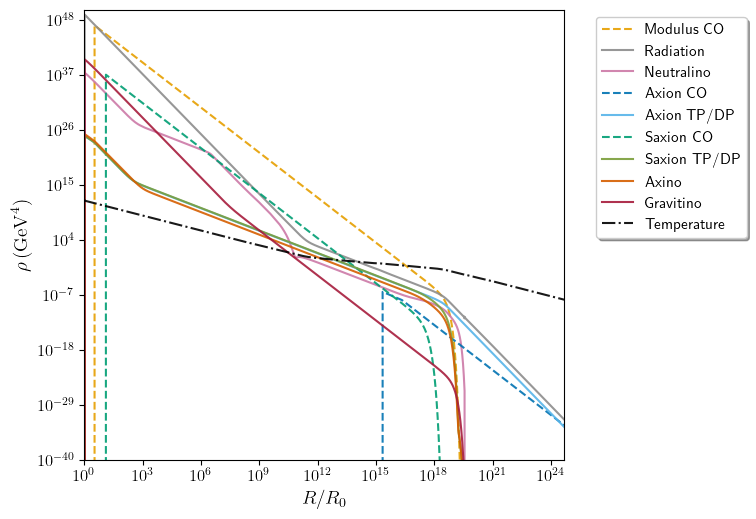}
      \caption{Plot of {\it a}) Yield $Y\equiv n/s$ and
        {\it b}) energy densities $\rho_i$ vs. scale factor
        $R/R_0$ for the PQMSSM with decaying neutralinos, leaving all axion DM which reproduces the observed abundance, $\Omega h^2 = 0.12$.
        We assume gravity-mediated SUSY breaking (SUGRA) with
        $f_a=1.5\times 10^{14}$ GeV, $m_{\phi} =200$ TeV, $m_{3/2}=100$ TeV,
        $m_s=m_{\ta}=30$ TeV, $\theta_i=2$, and $\lambda_{RPV}=10^{-7}$ for our
        standard natural SUSY BM model listed in the text.
        In {\it b}), we also display temperature (in units of GeV) for illustrative purposes of modulus decay.
     \label{fig:Yrho1}}
\end{figure}

We implement the decaying WIMP scenario into our nine-coupled Boltzmann treatment
for the evolution of various matter/radiation densities in the early universe
(for a detailed description of our coupled Boltzmann calculation,
see {\it e.g.} Ref. \cite{Baer:2023bbn}).
In Fig. \ref{fig:Yrho1}, we show in frame {\it a}) the evolution of the various
Yields ($Y\equiv n/s$) versus scale factor $R/R_0$, where $R_0$ is the scale factor 
at the time of reheat at the end of inflation.
Since the Yields scale the number density $n$ by the entropy density $s$, 
they are constant during periods of expansion when no new entropy is being created.
We see from the plot that at early times, the Yield is dominated by the 
CO-produced modulus field followed by the radiation density which is
generated by inflaton decay which is followed by the onset of oscillations for the CO-saxion. 
The TP/DP saxion and axino abundances climb with 
increasing $R/R_0$ as does the (pink) neutralino abundance until their production is sufficient to achieve an equilibrium density. 
The CO-produced axion abundance (dark-blue dashed curve) turns on around $R/R_0\sim 10^{15}$
and is then diluted due to modulus field decay (and also slightly by saxion, axino, and
neutralino decay) before it levels off after all decays have taken place. 
The neutralino abundance tries to freeze out, but is also diluted by various field
decays before decaying itself, leaving mainly axion CDM along with radiation, plus
a small amount of TP/DP axions.

In frame {\it b}), we show the energy densities $\rho_i$ of the various fields
(labeled by $i$) vs. $R/R_0$.
These distributions are all falling with increasing $R/R_0$ due to dilution
from the expanding universe.
We see that the early universe is very briefly radiation-dominated after
inflaton decay, but soon thereafter becomes modulus-dominated
(yellow-dashed curve) until the modulus field decays around $R/R_0\sim 10^{19}$. 
The cold neutralinos would quickly dominate over radiation once the modulus field decays, and be over-produced
(in accord with the WIMP-induced CMP) but instead they
(along with saxion, axino, and gravitino fields) all decay away around
$R/R_0\sim 10^{19}$ leaving radiation and TP/DP and CO-produced axions.
The radiation is diluted as $1/R^4$ while the cold CO-produced axions
dilute as $1/R^3$ and
so come to dominate the energy abundance around $R/R_0\sim 10^{25}$.
We also show the corresponding temperature $T$ by the black dot-dashed curve
which corresponds to the left-side vertical scale but in units of $T(GeV)$.

\subsection{Energy densities for Case 2}

\begin{figure}[htb!]
\centering
\includegraphics[height=0.4\textheight]{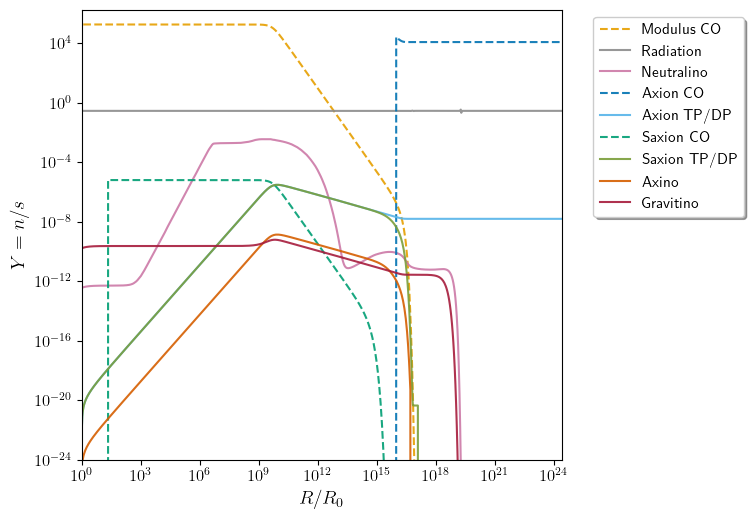}
\includegraphics[height=0.4\textheight]{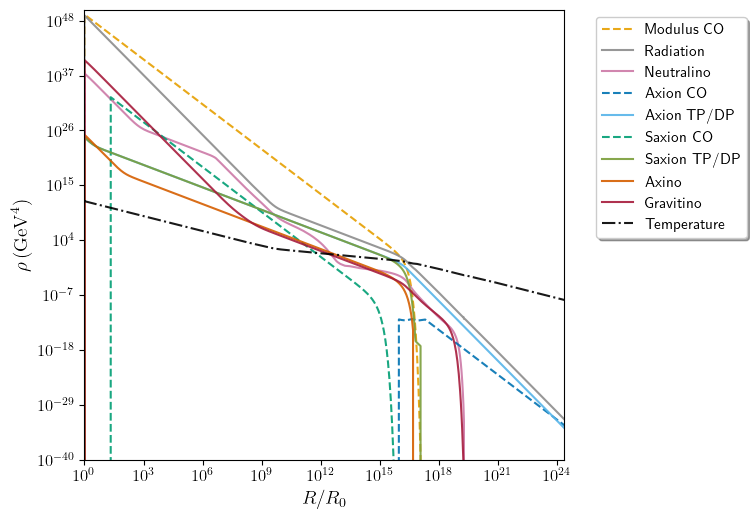}
      \caption{Plot of {\it a}) Yield $Y\equiv n/s$ and
        {\it b}) energy densities $\rho_i$ vs. scale factor
        $R/R_0$ for the PQMSSM with decaying neutralinos, leaving all axion DM which reproduces the observed abundance, $\Omega h^2 = 0.12$.
        We assume a KKLT-inspired set of scales with $m_{\phi}\sim 40 m_{3/2}$
      and take $f_a= 10^{12}$ GeV, $m_{\phi} =4$ PeV, $m_{3/2}=112$ TeV,
        $m_s=m_{\ta}=30$ TeV, $\theta_i=2$, and $\lambda_{RPV}=10^{-7}$ for our
        standard natural SUSY BM model listed in the text but for case 2:
        KKLT-inspired.
        In {\it b}), we also display temperature (in units of GeV) for illustrative purposes of modulus decay.
     \label{fig:Yrho2}}
\end{figure}

In Fig. \ref{fig:Yrho2}, we show the Yields $Y_i$ and the energy densities
$\rho_i$ vs. $R/R_0$ for the KKLT-inspired hierarchy of scales where
$m_{\phi}$ is allowed to be much heavier than $m_{3/2}$ and $m_{soft}$.
From frame {\it a}), we see that again the modulus field and radiation
early on will dominate the energy density of the universe, and since radiation
dilutes more quickly with the expansion, the modulus soon
dominates the energy density. The saxion begins decaying around
$R/R_0\sim 10^{15}$ thus feeding into the radiation abundance,
whose dilution visibly slows beginning around $R/R_0\sim 10^{10}$.
This is followed
by modulus, axino, and TP saxion decay around $R/R_0\sim 10^{17}$.
The gravitino and neutralino (dark and light purple curves) decay around
$R/R_0\sim 10^{19}$. With radiation and TP/DP axions diluting as $1/R^4$,
then cold, CO-produced axions become dominant around $R/R_0\agt 10^{24}$.

\subsection{Modulus decay through multiple long-lived particles}

\begin{figure}[htb!]
\centering
    {\includegraphics[height=0.3\textheight]{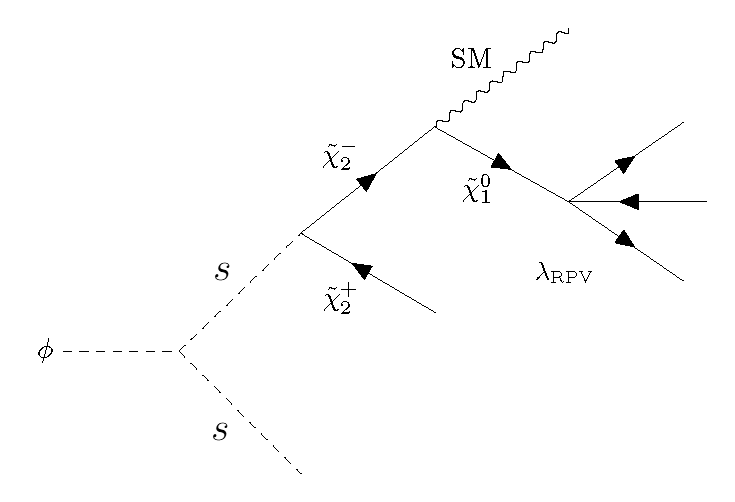}
      \caption{Cascade decay of long-lived modulus field $\phi$ via long-lived
        saxion $s$ and long-lived RPV neutralino $\tchi$.
        }
     \label{fig:phiss}}
\end{figure}

\begin{figure}[htb!]
\centering
    {\includegraphics[height=0.3\textheight]{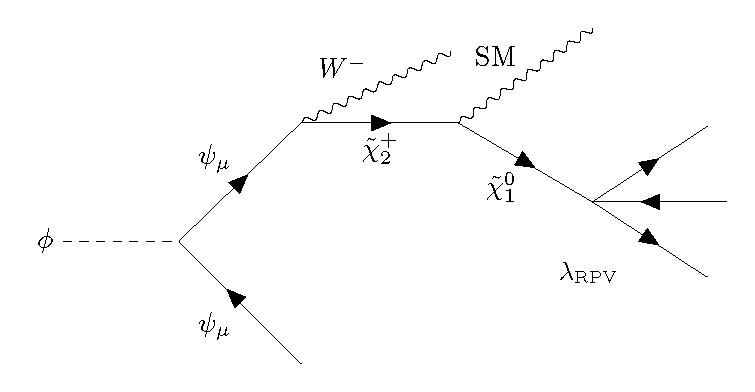}
      \caption{Cascade decay of long-lived modulus field $\phi$ via long-lived
        gravitino $\psi_\mu$ and long-lived RPV neutralino $\tchi$.
        }
     \label{fig:phiGG}}
\end{figure}

In the previous Section, we saw that the WIMP-induced CMP is greatly
ameliorated in the decaying WIMP scenarios since all the WIMP DM decays away
before BBN, leaving only axion dark matter which is primarily cold.
However, an old issue arises with a new slant.
The long-lived modulus field must of course obey BBN constraints on long-lived
neutral relics in the early universe\cite{Kawasaki:2004qu,Jedamzik:2006xz,Kawasaki:2008qe}, 
but now the possibility arises of {\it modulus cascade decays} through more
than one step of long-lived particles.
For instance, the modulus field $\phi$ decays also act as a source for
long-lived saxions, axinos, neutralinos, and gravitinos.
These decays may occur in several steps: {\it e.g.} 
$\phi\to s,\ta \to\tchi_1^0\to SM$ particles and the gravitino may enter
the decay chain as well if decays to $\psi_\mu$ are allowed: see
{\it e.g.} Figs. \ref{fig:phiss} and \ref{fig:phiGG}.

For the $\tchi_1^0\to ff^\prime f^{\prime\prime}$ decay, we adopt the approximate
formula for photino decay\cite{Dreiner:1997uz}
\be
\Gamma (\tchi_1^0 )\simeq n_{modes}\frac{3\alpha\lambda_{RPV}^2}{128\pi^2}
\frac{m_{\tchi_1^0}^5}{m_{soft}^4}
\ee
where  $n_{modes}$ counts the number of RPV decay modes of the $\tchi_1^0$
(the various possibilities are tabulated in Ref. \cite{Baer:2025srs}; here for
simplicity we absorb $n_{modes}$ into our definition of $\lambda_{RPV}^2$, which provides a small amount of additional freedom in our assumed values of $\lambda_{RPV}$).
This approximate formula gives accuracy to a factor of several against
exact tree-level calculations\cite{Baer:2025srs}.

To handle this case of multi-step long-lived particle (LLP) decays, we simply
sum over the (relativistically corrected) lifetimes of the LLPs in the
multistep cascade decay to estimate the mean total lifetime.
For instance, for the decay chain $A\to B\to X$, we have
\be
\tau_A^{tot}\simeq \tau (A\to B) +\gamma_B\tau (B\to X)
\ee
where $\gamma_B=E_B/m_B$ is the relativistic time dilation factor.
Throughout this work, we use the numerical solutions of the Boltzmann equations to determine the $\gamma_B$ factors, where $E_B$ is evaluated at the time when $3H = \Gamma_A$ (i.e. the time of $B$'s production).
The relativistic correction can be significant in that the daughter particles
produced in the heavy particle decay can be produced with $E\gg m$.
Although this estimate neglects the expansion of the Universe during the cascade decay, 
as well as the daughter-particle energy distributions associated with $N$-body decays, 
it provides a suitable benchmark for estimating the characteristic timescale relevant for BBN constraints.

The two-body modulus decay widths have been calculated in
Ref. \cite{Bae:2022okh} where we assume the case of helicity-suppressed gravitinos (i.e. $\Gamma_{\phi \, \rightarrow \, \psi_\mu \psi_\mu} \sim m_\phi m_{3/2}^2/m_P^2$), and the saxion and axino decay widths are
from Ref. \cite{Bae:2013hma}
(for the case of the SUSY DFSZ axion model).
The gravitino decay widths are adopted from Ref. \cite{Kohri:2005wn}.

\begin{figure}[htb!]
\centering
\includegraphics[height=0.4\textheight]{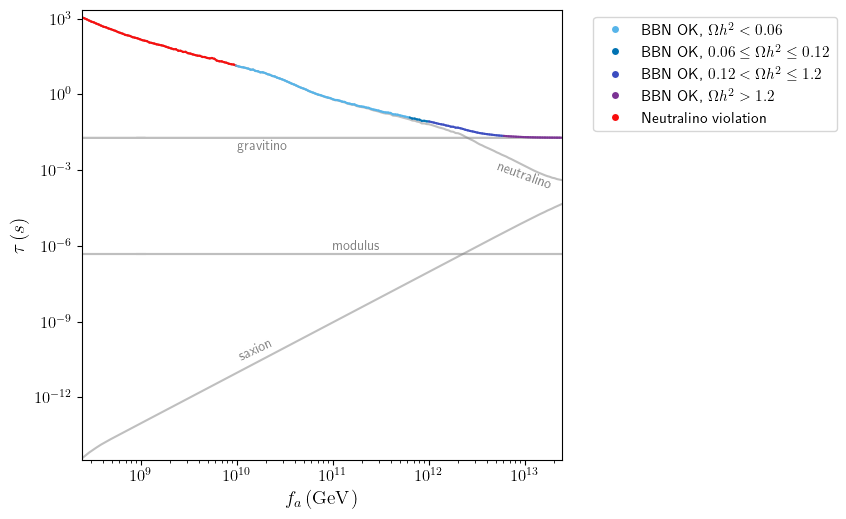}\\
\includegraphics[height=0.4\textheight]{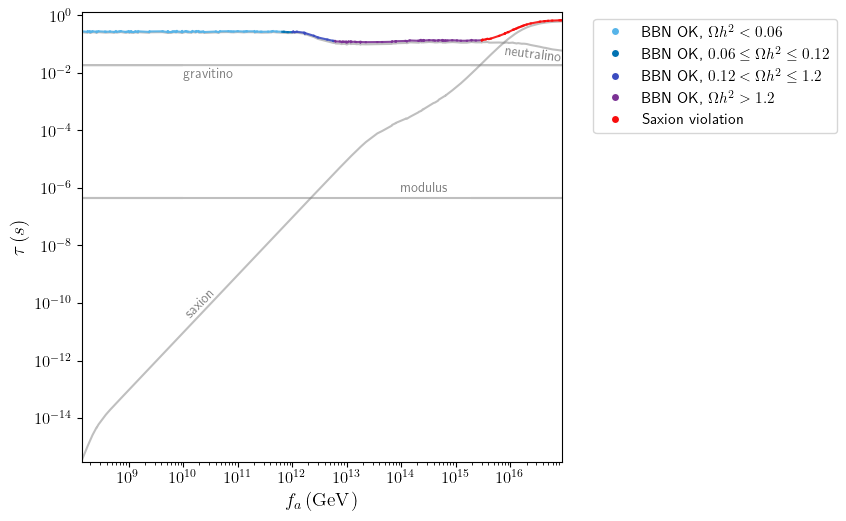}\\
\caption{Plot of lifetimes $\tau$ vs. $f_a$  for the KKLT-inspired
  scale choices with $m_{\phi}=4$ PeV, $m_{3/2}=100$ TeV and our usual SUSY
  BM point. 
  In {\it a}), we show lifetimes where $\lambda_{RPV}\sim (f_a/m_P)^1$
  while in {\it b}) we take $\lambda_{RPV}$ fixed at $10^{-7}$.
  The colored curve estimates the total lifetime of the cascade decay $\phi \rightarrow $ final state.
\label{fig:tau}}
\end{figure}

To illustrate, in Fig. \ref{fig:tau} we show the lifetime $\tau$ of 
various of the LLPs that occur in the PQMSSM for our BM scenario
but versus PQ scale $f_a$ and for the KKLT case with $m_{\phi}=4$ PeV.
In each frame, the colored curve displays the estimated total lifetime of the cascade decay beginning with the modulus and (eventually) ending with neutralino decay.
In frame {\it a}), we show the lifetimes for the case where
$\lambda_{RPV}\sim (f_a/m_P)^1$. 
In this case, the modulus and gravitino lifetimes are fixed as $f_a$ increases, 
but the saxion lifetime increases with increasing $f_a$
since its width $\Gamma_s\sim m_s^3/f_a^2$.
In contrast, the neutralino width goes as
$\Gamma_{\tchi}\sim \lambda_{RPV}^2\sim f_a^2$ so that the lifetime $\tau (\tchi )$ decreases with increasing $f_a$. 
The red-shaded portion of the $\tau (\tchi )$ curve is ruled out by
neutralino disruption of successful BBN\cite{Jedamzik:2006xz} for
$f_a\alt 10^{10}$ GeV. 
For larger $f_a$ values, $\tchi$ decays
before the onset of BBN -- however, the remaining axion DM abundance is still subjected to entropy dilution from the late-decaying modulus, resulting in an underproduced total DM abundance shown by the light blue-shaded portion of the curve.
As the axion relic density behaves as $\Omega_a^{CO}h^2\sim f_a^{7/6}$, 
the entropy dilution from the modulus is partially offset, resulting in a viable window of all-axion DM (dark blue-shaded portion).
Eventually, too much axion DM is produced for $f_a\agt 5\times 10^{12}$ GeV, as shown by the purple-shaded portion of the curve.

In frame {\it b}), we show a similar plot, but this time instead
keeping a fixed value of $\lambda_{RPV}=10^{-7}$. In this case, the region
with low $f_a$ is no longer excluded by BBN bounds. 
Additionally, we see that near $f_a \sim 10^{12}$ GeV, the neutralino lifetime decreases once the saxion becomes longer-lived than the modulus.
In this case, the relativistic correction for the neutralino becomes much smaller given the assumed mass hierarchy in the figure, $m_\phi \sim \mathcal{O}(10^2) m_s \sim \mathcal{O}(10^4) m_{\tchi_1^0}$.
Once $f_a\agt 5\times 10^{15}$ GeV,
the model again violates BBN bounds due to the late decay of
saxions, which also lengthens the scale of the neutralino decay.
The axion DM abundance is essentially unaffected by this change to $\lambda_{RPV}$.

\section{Results in $m_{\phi}$ vs. $m_{3/2}$ plane}

In this Section, we fold all our results together for the case of SUSY
with  decaying WIMPs to show the ultimate bounds in
the modulus mass $m_\phi$ vs. gravitino mass $m_{3/2}$
parameter space from the CMP.

\begin{figure}[htb!]
\centering
\includegraphics[height=0.4\textheight]{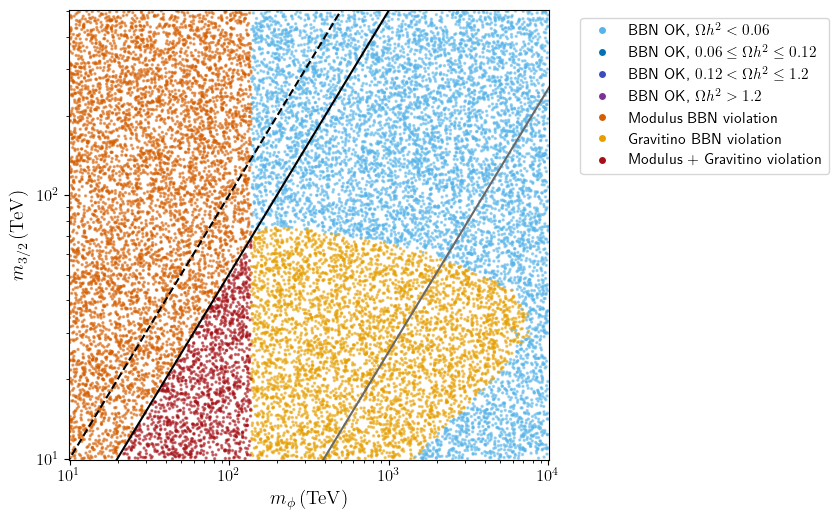}\\
\includegraphics[height=0.3\textheight]{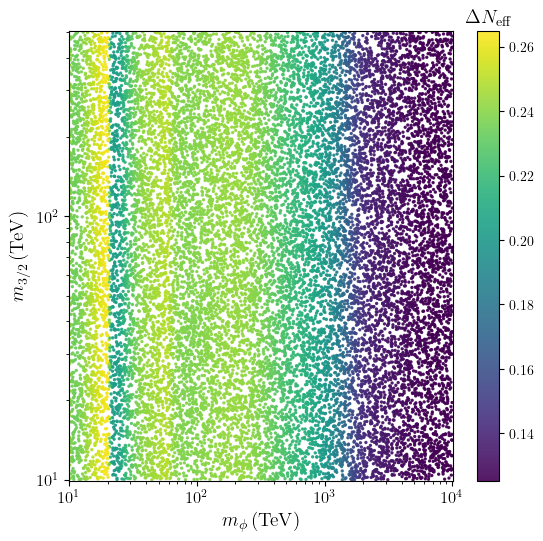}
\includegraphics[height=0.3\textheight]{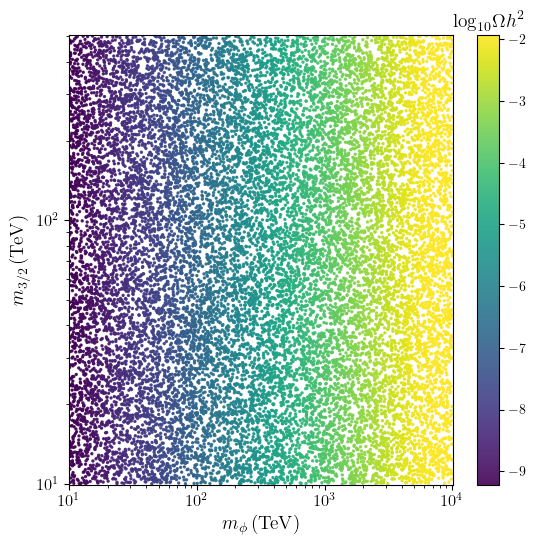}
      \caption{Allowed/restricted regions of $m_{\phi}$ vs. $m_{3/2}$ plane
        for $\theta_i=2$, $f_a=10^{11}$ GeV, and $\lambda_{RPV}=10^{-7} \simeq f_a/m_P$.
        In the upper plot, orange is BBN-violating for $\phi$ only, yellow is BBN-violating for
        $\psi_\mu$-only, red is BBN-violating for both $\phi$ and $\psi_\mu$, light blue is BBN-allowed with $\Omega_a h^2<0.06$, medium blue has
        $0.06 <\Omega_a h^2 < 0.12$, dark blue has $0.12 < \Omega_a h^2 < 1.2$, and purple has $\Omega_a h^2 > 1.2$.
     \label{fig:plane11}}
\end{figure}

In Fig. \ref{fig:plane11}, we show the $m_\phi$ vs. $m_{3/2}$ plane for our
natSUSY BM point and with PQ parameters $\theta_i =2$, $f_a=10^{11}$ GeV, and
$\lambda_{RPV}=10^{-7}$.
Frame {\it a}) first shows BBN-allowed and BBN-forbidden regions of the
plane.
The dashed black line shows where $m_\phi =m_{3/2}$, to guide the reader.
Below the solid black line is where the $\phi\to \psi_\mu \psi_\mu$ decay is open.
Thus, the region between the black lines is roughly favored by the SUGRA
mass scales with $m_\phi\sim m_{3/2}$.
The solid gray line shows the KKLT-inspired region where $m_\phi\sim 40m_{3/2}$.

The orange-colored dots denote BBN violating $\phi$ decays, the traditional
bound for the CMP, where $m_\phi\agt 130$ TeV is required.
The yellow-colored dots denote the region which is excluded by late-decaying
gravitinos, both thermally and $\phi$-decay produced.
The red-shaded region is excluded both by late decaying moduli {\it and}
late-decaying gravitinos. 
Meanwhile, the blue shaded regions are BBN allowed, with light blue having 
$\Omega_ah^2\alt 0.06$; for plots with larger $f_a$ values,
then medium-blue has $0.06<\Omega_ah^2<0.12$ and darker blue has
$1.2 > \Omega_ah^2>0.12$ while purple has $\Omega_ah^2>1.2$.
The CO-produced axion abundance can be raised or lowered by
changing the value of the misalignment angle $\theta_i$.
In the case shown in Fig. \ref{fig:plane11}, the axion DM abundance is always below the measured
value, due to the substantial entropy dilution from modulus decay.

From the plot, we see that generally gravitinos with mass $m_{3/2}\agt 70$ TeV
are required to be BBN safe.
This is considerably higher than older bounds\cite{Kawasaki:2008qe}
where $m_{3/2}\agt 5-10$ TeV is required. 
The higher mass bounds are due to the fact that here modulus decay is now
acting as a substantial source for gravitinos,
requiring the population to decay entirely prior to BBN rather than
retaining a sufficiently small abundance at the onset of BBN.
However, as we are assuming the case of helicity-suppressed modulus decays
to gravitinos ($\Gamma_{\phi \, \rightarrow \, \psi_\mu \psi_\mu} \sim m_\phi m_{3/2}^2/m_P^2$), the relevant branching fraction can become small enough to bring the abundance of lighter gravitinos into accord with BBN constraints once $m_\phi\gtrsim 10^3$ TeV\cite{Bae:2022okh}.
For the case of unsuppressed modulus decays to gravitinos ($\Gamma_{\phi \, \rightarrow \, \psi_\mu \psi_\mu} \sim m_\phi^3/m_P^2$),
this small region allowing light gravitinos would disappear.
From the plot, we also see that $m_\phi\agt 130$ TeV is required.
This is well below the value derived from the WIMP-induced CMP where
$m_\phi\agt 5-10$ PeV is required\cite{Bae:2022okh,Baer:2023bbn}.
Notably, this is also higher than the typically-quoted BBN
bound of $m_\phi\agt 100$ TeV, which is computed by calculating the modulus
decay temperature 
$
  T_D^\phi 
  \sim
  \sqrt{
    m_P
    \Gamma_\phi
  }
$ and setting this to the BBN temperature, $T_{BBN} \sim 3-5$ MeV.
In this work, our results instead are computed from the bounds on abundances of decaying neutral particles presented in \cite{Jedamzik:2006xz}, so that we instead verify whether the energy density of the modulus at the time of decay, along with its lifetime and hadronic branching fraction, lead to unacceptable production or disruption of light nuclei.

In Fig. \ref{fig:plane11} frame {\it b}), we show the amount of dark
radiation produced over the $m_\phi$ vs. $m_{3/2}$ plane, as encoded in its contribution to the
effective number of neutrinos $\Delta N_{\text{eff}}$.
The measured value from the Particle Data Group (PDG)\cite{ParticleData:2026uvj}
(from Planck CMB plus BAO data) is $N_{\text{eff}}=2.99\pm 0.17$
while the SM predicts $N_{\text{eff}}=3.044$ (thus, there is actually a slight
dearth in the measured value as compared to SM predictions).
The contribution from the $\phi$PQMSSM model
with decaying neutralinos is that $\Delta N_{\text{eff}}\sim 0.14-0.26$ coming
primarily from decay-produced axions.
Thus, the data favor the largest values of $m_\phi$ with lower
$\Delta N_{\text{eff}}$ values.

In frame {\it c}), we show the CO-produced axion dark matter relic density $\Omega_ah^2$
as computed by our nine-coupled Boltzmann equation code.
The values range from $\Omega_ah^2\sim 10^{-9}$ on the left edge to
$\sim 10^{-2}$ on the right edge. These very low values of all-axion DM
reflect the substantial entropy dilution of the CO-produced axions
by the late-decaying modulus.
One may increase the relic abundance as usual by increasing
the axion field misalignment angle $\theta_i\to \pi$.

\begin{figure}[htb!]
\centering
\includegraphics[height=0.4\textheight]{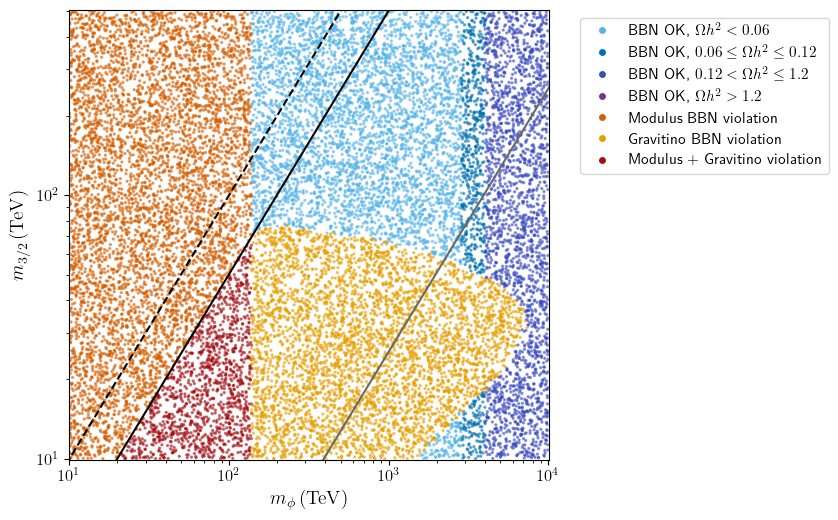}\\
\includegraphics[height=0.3\textheight]{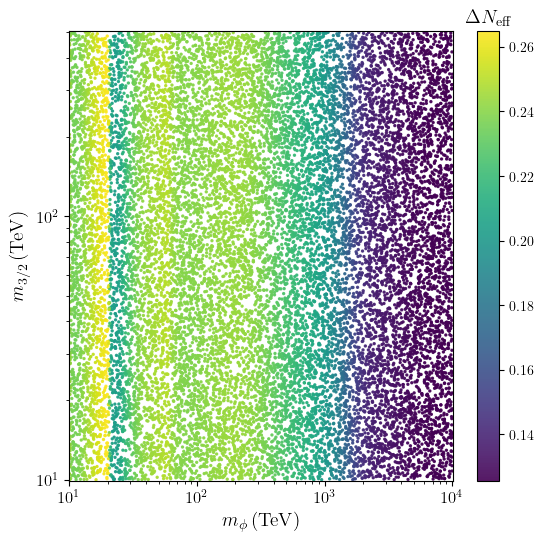}
\includegraphics[height=0.3\textheight]{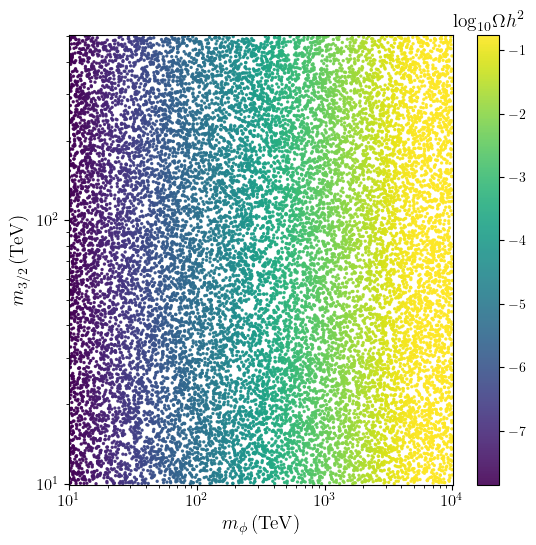}
      \caption{Allowed/restricted regions of $m_{\phi}$ vs. $m_{3/2}$ plane
        for $\theta_i=2$, $f_a=10^{12}$ GeV, and $\lambda_{RPV}=10^{-7}$.
        In the upper plot, orange is BBN-violating for $\phi$ only, yellow is BBN-violating for
        $\psi_\mu$-only, red is BBN-violating for both $\phi$ and $\psi_\mu$, light blue is BBN-allowed with $\Omega_a h^2<0.06$, medium blue has
        $0.06 <\Omega_a h^2 < 0.12$, dark blue has $0.12 < \Omega_a h^2 < 1.2$, and purple has $\Omega_a h^2 > 1.2$.
     \label{fig:plane12}}
\end{figure}

In Fig. \ref{fig:plane12}, we show the same $m_\phi$ vs. $m_{3/2}$
set of planes but instead take $f_a=10^{12}$ GeV.
From frame {\it a}), we see that the BBN-constrained region is similar
to that shown in Fig. \ref{fig:plane11}.
What mainly changes now is the expected axion DM abundance over the plot.
With $f_a=10^{12}$ GeV, we gain more CO-produced axions, which (partially) offsets the entropy dilution from the modulus.
Thus, in frame {\it a}), we see patches of medium-blue and dark blue
on the right-hand-side of the plot corresponding to $\Omega_ah^2\sim 0.12$.
These regions correspond to $m_\phi\sim 3-10$ PeV.
This is also shown in frame {\it c}) where the right-hand-side of the
plot has $\Omega_ah^2\sim 0.12$.
In this region, one can start to reconcile naturalness with the CMP in
the KKLT-inspired mass scale case where
$m_{soft}\sim m_{3/2}/40\sim m_\phi/(40)^2$.

\begin{figure}[htb!]
\centering
\includegraphics[height=0.4\textheight]{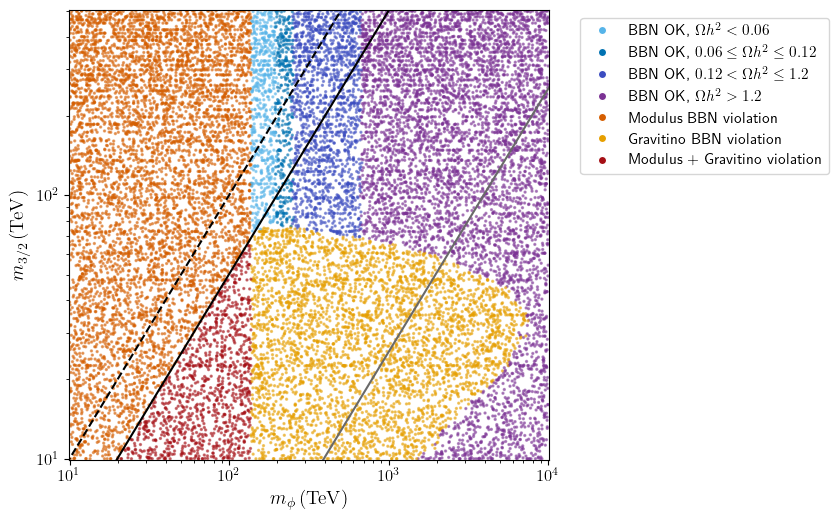}\\
\includegraphics[height=0.3\textheight]{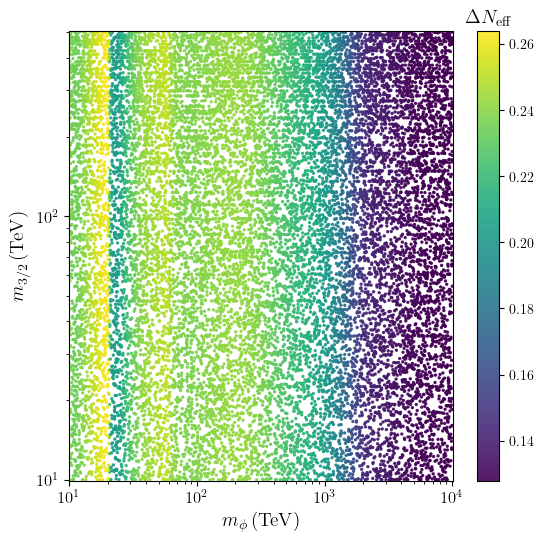}
\includegraphics[height=0.3\textheight]{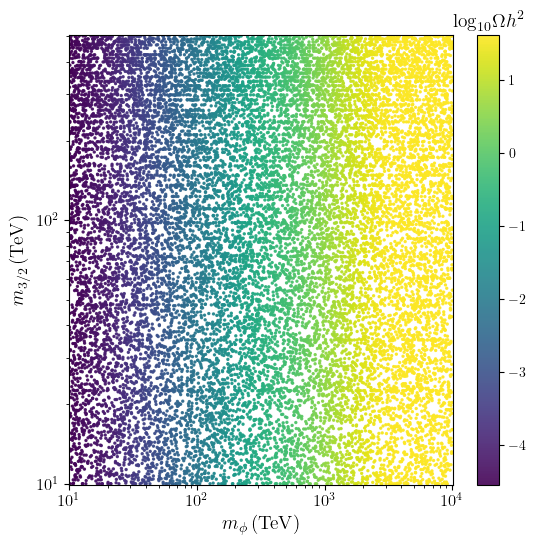}
      \caption{Allowed/restricted regions of $m_{\phi}$ vs. $m_{3/2}$ plane
        for $\theta_i=2$, $f_a=10^{14}$ GeV, and $\lambda_{RPV}=10^{-7}$.
        In the upper plot, orange is BBN-violating for $\phi$ only, yellow is BBN-violating for
        $\psi_\mu$-only, red is BBN-violating for both $\phi$ and $\psi_\mu$, light blue is BBN-allowed with $\Omega_a h^2<0.06$, medium blue has
        $0.06 <\Omega_a h^2 < 0.12$, dark blue has $0.12 < \Omega_a h^2 < 1.2$, and purple has $\Omega_a h^2 > 1.2$.
     \label{fig:plane14}}
\end{figure}

In Fig. \ref{fig:plane14}, we again show the $m_\phi$ vs. $m_{3/2}$ set of plane
plots, but now for $f_a=10^{14}$ GeV. Here, the BBN and $\Delta N_{\text{eff}}$
regions are roughly the same as for the lower $f_a$ cases, but the axion-only relic density has increased
much further, offsetting much of the entropy dilution of the CO-produced axion relic density.
From frames {\it a}) and {\it c}), we see that $\Omega_ah^2\sim 0.12$
for $m_\phi\sim 200$ TeV. Such a low value of $m_\phi$ means $m_{3/2}\sim m_\phi$,
so now we are in the SUGRA-preferred mass range, and very far from the
gray KKLT line, where axion dark matter is greatly over produced.
%

\section{Conclusions}
\label{sec:conclude}

In this paper, we have examined the impact of decaying neutralinos on the
cosmological moduli problem. In the case of stable neutralino plus axion
mixed dark matter, values of at least $m_\phi\sim 5-10$ PeV are required
to avoid overproduction of WIMP dark matter arising from moduli-field decays.
Our new considerations presented here pertain to at least two models of
decaying neutralinos.
1. One possibility is a model where the axino is instead the LSP, so
$\tchi_1^0\to \ta+(Z,h)$.
Then the dark matter can be either {\it a}) mostly warm axinos plus some
axions or {\it b}) mostly cold axions with a smattering of axinos.
2. A second possibility comes from a model where discrete $\mathbb{Z}_n^R$
symmetries  (which can arise from string compactifications) are used to
suppress the SUSY $\mu$ term, and also generates an accidental, approximate
global $U(1)_{PQ}$ symmetry needed to solve the strong CP problem. 
Here, RPV interactions are generated from higher-dimensional operators arising
from the new PQ charged field coupling to visible sector superfields,
where the RPV couplings are suppressed as $(f_a/m_P)^n$.
For the cases where $n=1$, then all neutralinos decay in the
early universe before the onset of BBN, leading to all-axion dark matter
from SUSY.

For both these cases, the CMP is greatly ameliorated since the moduli-induced
WIMP LSP problem, which places a more severe constraint on $m_\phi$ than
does BBN, is now avoided.
Previously, this required a lightest modulus
mass typically $m_\phi \agt 5-10$ PeV, creating strong tension with naturalness
in the case where the modulus mass and visible sector soft terms are comparable.
While DM overproduction in the decaying WIMP scenarios is overcome,
the CMP-induced BBN problem can be enhanced because now one must combine
lifetimes of the decaying modulus with the decaying WIMP (and possibly other
LLPs such as saxion or gravitino), and require the sum of the lifetimes along the decay chain to be shorter than $t_{BBN}$.
In the KKLT-like case where
$m_\phi\gg m_{3/2}\gg m_{soft}$, then one gets three stage delayed modulus decays
such as $\phi\to\psi_\mu\to \tchi_1^0 $ so one must require
the sum $\tau_\phi +\tau_{3/2}+\tau_{\tchi}$ to be less than $t_{BBN}$.
These considerations  typically require $m_{\phi}\agt 130$ TeV which is a
big improvement over the multi-PeV bound from the moduli-induced DM
overproduction problem. The new result is much easier to reconcile with
EW naturalness in the SUSY spectrum, especially in the case where a KKLT-like 
hierarchy of scales is present.

{\it Acknowledgements:} 
HB gratefully acknowledges support from the Avenir Foundation.
VB gratefully acknowledges support from the U.S. Department of Energy,
Office of Science, Office of High Energy Physics,
under Award Number DE-SC0017647 and from the William F. Vilas Estate.


\bibliography{mod3}

@article{Dreiner:1997uz,
    author = "Dreiner, Herbert K.",
    editor = "Kane, Gordon L.",
    title = "{An Introduction to explicit R-parity violation}",
    eprint = "hep-ph/9707435",
    archivePrefix = "arXiv",
    doi = "10.1142/9789814307505_0017",
    journal = "Adv. Ser. Direct. High Energy Phys.",
    volume = "21",
    pages = "565--583",
    year = "2010"
}

@article{Kawasaki:2004qu,
    author = "Kawasaki, Masahiro and Kohri, Kazunori and Moroi, Takeo",
    title = "{Big-Bang nucleosynthesis and hadronic decay of long-lived massive particles}",
    eprint = "astro-ph/0408426",
    archivePrefix = "arXiv",
    reportNumber = "ICRR-REPORT-508-2004-6, OU-TAP-234, TU-727",
    doi = "10.1103/PhysRevD.71.083502",
    journal = "Phys. Rev. D",
    volume = "71",
    pages = "083502",
    year = "2005"
}

@article{Jedamzik:2006xz,
    author = "Jedamzik, Karsten",
    title = "{Big bang nucleosynthesis constraints on hadronically and electromagnetically decaying relic neutral particles}",
    eprint = "hep-ph/0604251",
    archivePrefix = "arXiv",
    doi = "10.1103/PhysRevD.74.103509",
    journal = "Phys. Rev. D",
    volume = "74",
    pages = "103509",
    year = "2006"
}

@article{Baer:2026jeb,
    author = "Baer, Howard and Barger, Vernon and Sengupta, Dibyashree",
    title = "{Serendipitous supersymmetric solution to the strong CP problem}",
    eprint = "2607.07921",
    archivePrefix = "arXiv",
    primaryClass = "hep-ph",
    reportNumber = "OU-HEP-260703",
    month = "7",
    year = "2026"
}

@article{ParticleData:2026uvj,
    author = "Takahashi, F. and others",
    collaboration = "Particle Data",
    title = "{Review of Particle Physics*}",
    doi = "10.1142/s0217751x26300115",
    journal = "Int. J. Mod. Phys. A",
    volume = "41",
    number = "22",
    pages = "2630011",
    year = "2026"
}

@article{Nilles:2001ap,
    author = "Nilles, Hans Peter",
    editor = "Olive, K. A. and Rudaz, S. and Shifman, Mikhail A.",
    title = "{Hidden sector supergravity breakdown}",
    eprint = "hep-ph/0106063",
    archivePrefix = "arXiv",
    doi = "10.1016/S0920-5632(01)01508-0",
    journal = "Nucl. Phys. B Proc. Suppl.",
    volume = "101",
    pages = "237--250",
    year = "2001"
}

@article{Jungman:1995df,
    author = "Jungman, Gerard and Kamionkowski, Marc and Griest, Kim",
    title = "{Supersymmetric dark matter}",
    eprint = "hep-ph/9506380",
    archivePrefix = "arXiv",
    reportNumber = "SU-4240-605, UCSD-PTH-95-02, IASSNS-HEP-95-14, CU-TP-677",
    doi = "10.1016/0370-1573(95)00058-5",
    journal = "Phys. Rept.",
    volume = "267",
    pages = "195--373",
    year = "1996"
}

@article{Bailin:1987jd,
    author = "Bailin, D. and Love, A.",
    title = "{KALUZA-KLEIN THEORIES}",
    doi = "10.1088/0034-4885/50/9/001",
    journal = "Rept. Prog. Phys.",
    volume = "50",
    pages = "1087--1170",
    year = "1987"
}

@article{Horava:1996ma,
    author = "Horava, Petr and Witten, Edward",
    title = "{Eleven-dimensional supergravity on a manifold with boundary}",
    eprint = "hep-th/9603142",
    archivePrefix = "arXiv",
    reportNumber = "IASSNS-HEP-96-17, PUPT-1597",
    doi = "10.1016/0550-3213(96)00308-2",
    journal = "Nucl. Phys. B",
    volume = "475",
    pages = "94--114",
    year = "1996"
}

@article{Candelas:1985en,
    author = "Candelas, P. and Horowitz, Gary T. and Strominger, Andrew and Witten, Edward",
    title = "{Vacuum configurations for superstrings}",
    reportNumber = "NSF-ITP-84-170",
    doi = "10.1016/0550-3213(85)90602-9",
    journal = "Nucl. Phys. B",
    volume = "258",
    pages = "46--74",
    year = "1985"
}

@article{Kachru:2003aw,
    author = "Kachru, Shamit and Kallosh, Renata and Linde, Andrei D. and Trivedi, Sandip P.",
    title = "{De Sitter vacua in string theory}",
    eprint = "hep-th/0301240",
    archivePrefix = "arXiv",
    reportNumber = "SLAC-PUB-9630, SU-ITP-03-01, TIFR-TH-03-03",
    doi = "10.1103/PhysRevD.68.046005",
    journal = "Phys. Rev. D",
    volume = "68",
    pages = "046005",
    year = "2003"
}

@article{Acharya:2010af,
    author = "Acharya, Bobby Samir and Kane, Gordon and Kuflik, Eric",
    title = "{Bounds on scalar masses in theories of moduli stabilization}",
    eprint = "1006.3272",
    archivePrefix = "arXiv",
    primaryClass = "hep-ph",
    reportNumber = "MCTP-10-48",
    doi = "10.1142/S0217751X14500730",
    journal = "Int. J. Mod. Phys. A",
    volume = "29",
    pages = "1450073",
    year = "2014"
}

@article{Kane:2015jia,
    author = "Kane, Gordon and Sinha, Kuver and Watson, Scott",
    title = "{Cosmological Moduli and the Post-Inflationary Universe: A Critical Review}",
    eprint = "1502.07746",
    archivePrefix = "arXiv",
    primaryClass = "hep-th",
    doi = "10.1142/S0218271815300220",
    journal = "Int. J. Mod. Phys. D",
    volume = "24",
    number = "08",
    pages = "1530022",
    year = "2015"
}

@article{Coughlan:1983ci,
    author = "Coughlan, G. D. and Fischler, W. and Kolb, Edward W. and Raby, S. and Ross, Graham G.",
    title = "{Cosmological Problems for the Polonyi Potential}",
    reportNumber = "LA-UR-83-1423",
    doi = "10.1016/0370-2693(83)91091-2",
    journal = "Phys. Lett. B",
    volume = "131",
    pages = "59--64",
    year = "1983"
}

@article{deCarlos:1993wie,
    author = "de Carlos, B. and Casas, J. A. and Quevedo, F. and Roulet, E.",
    title = "{Model independent properties and cosmological implications of the dilaton and moduli sectors of 4-d strings}",
    eprint = "hep-ph/9308325",
    archivePrefix = "arXiv",
    reportNumber = "CERN-TH-6958-93, NEIP-93-006, IEM-FT-75-93",
    doi = "10.1016/0370-2693(93)91538-X",
    journal = "Phys. Lett. B",
    volume = "318",
    pages = "447--456",
    year = "1993"
}

@article{Banks:1993en,
    author = "Banks, Tom and Kaplan, David B. and Nelson, Ann E.",
    title = "{Cosmological implications of dynamical supersymmetry breaking}",
    eprint = "hep-ph/9308292",
    archivePrefix = "arXiv",
    reportNumber = "UCSD-PTH-93-26, RU-37",
    doi = "10.1103/PhysRevD.49.779",
    journal = "Phys. Rev. D",
    volume = "49",
    pages = "779--787",
    year = "1994"
}

@article{Planck:2018vyg,
    author = "Aghanim, N. and others",
    collaboration = "Planck",
    title = "{Planck 2018 results. VI. Cosmological parameters}",
    eprint = "1807.06209",
    archivePrefix = "arXiv",
    primaryClass = "astro-ph.CO",
    doi = "10.1051/0004-6361/201833910",
    journal = "Astron. Astrophys.",
    volume = "641",
    pages = "A6",
    year = "2020",
    note = "[Erratum: Astron.Astrophys. 652, C4 (2021)]"
}

@article{Nilles:2013lda,
    author = "Nilles, Hans Peter and Ramos-S{\'a}nchez, Sa{\'u}l and Ratz, Michael and Vaudrevange, Patrick K. S.",
    title = "{A note on discrete $R$ symmetries in $\mathbb{Z}_{6}$-II orbifolds with Wilson lines}",
    eprint = "1308.3435",
    archivePrefix = "arXiv",
    primaryClass = "hep-th",
    reportNumber = "DESY-13-143, TUM-HEP-901-13, FLAVOUR-EU-52-13",
    doi = "10.1016/j.physletb.2013.09.041",
    journal = "Phys. Lett. B",
    volume = "726",
    pages = "876--881",
    year = "2013"
}

@article{Conlon:2015uua,
    author = "Conlon, Joseph",
    title = "{The What and Why of Moduli}",
    doi = "10.1142/9789814602686_0002",
    journal = "Adv. Ser. Direct. High Energy Phys.",
    volume = "22",
    pages = "11--22",
    year = "2015"
}

@article{Kawasaki:1995cy,
    author = "Kawasaki, M. and Moroi, T. and Yanagida, T.",
    title = "{Constraint on the reheating temperature from the decay of the Polonyi field}",
    eprint = "hep-ph/9509399",
    archivePrefix = "arXiv",
    reportNumber = "ICRR-340-95-6, LBL-37715, UT-719",
    doi = "10.1016/0370-2693(95)01546-9",
    journal = "Phys. Lett. B",
    volume = "370",
    pages = "52--58",
    year = "1996"
}

@article{Baer:2012up,
    author = "Baer, Howard and Barger, Vernon and Huang, Peisi and Mustafayev, Azar and Tata, Xerxes",
    title = "{Radiative natural SUSY with a 125 GeV Higgs boson}",
    eprint = "1207.3343",
    archivePrefix = "arXiv",
    primaryClass = "hep-ph",
    doi = "10.1103/PhysRevLett.109.161802",
    journal = "Phys. Rev. Lett.",
    volume = "109",
    pages = "161802",
    year = "2012"
}

@article{Baer:2012cf,
    author = "Baer, Howard and Barger, Vernon and Huang, Peisi and Mickelson, Dan and Mustafayev, Azar and Tata, Xerxes",
    title = "{Radiative natural supersymmetry: Reconciling electroweak fine-tuning and the Higgs boson mass}",
    eprint = "1212.2655",
    archivePrefix = "arXiv",
    primaryClass = "hep-ph",
    doi = "10.1103/PhysRevD.87.115028",
    journal = "Phys. Rev. D",
    volume = "87",
    number = "11",
    pages = "115028",
    year = "2013"
}

@article{Choi:2005ge,
    author = "Choi, Kiwoon and Falkowski, Adam and Nilles, Hans Peter and Olechowski, Marek",
    title = "{Soft supersymmetry breaking in KKLT flux compactification}",
    eprint = "hep-th/0503216",
    archivePrefix = "arXiv",
    reportNumber = "IFT-05-05, KAIST-TH-2005-04, DESY-05-052",
    doi = "10.1016/j.nuclphysb.2005.04.032",
    journal = "Nucl. Phys. B",
    volume = "718",
    pages = "113--133",
    year = "2005"
}

@article{Choi:2004sx,
    author = "Choi, K. and Falkowski, A. and Nilles, Hans Peter and Olechowski, M. and Pokorski, S.",
    title = "{Stability of flux compactifications and the pattern of supersymmetry breaking}",
    eprint = "hep-th/0411066",
    archivePrefix = "arXiv",
    reportNumber = "IFT-29-2004, KAIST-TH-2004-19",
    doi = "10.1088/1126-6708/2004/11/076",
    journal = "JHEP",
    volume = "11",
    pages = "076",
    year = "2004"
}

@article{Kawasaki:2006gs,
    author = "Kawasaki, Masahiro and Takahashi, Fuminobu and Yanagida, T. T.",
    title = "{Gravitino overproduction in inflaton decay}",
    eprint = "hep-ph/0603265",
    archivePrefix = "arXiv",
    reportNumber = "DESY-06-034",
    doi = "10.1016/j.physletb.2006.05.037",
    journal = "Phys. Lett. B",
    volume = "638",
    pages = "8--12",
    year = "2006"
}

@article{Nakamura:2006uc,
    author = "Nakamura, Shuntaro and Yamaguchi, Masahiro",
    title = "{Gravitino production from heavy moduli decay and cosmological moduli problem revived}",
    eprint = "hep-ph/0602081",
    archivePrefix = "arXiv",
    reportNumber = "TU-765",
    doi = "10.1016/j.physletb.2006.05.078",
    journal = "Phys. Lett. B",
    volume = "638",
    pages = "389--395",
    year = "2006"
}

@article{Endo:2006zj,
    author = "Endo, Motoi and Hamaguchi, Koichi and Takahashi, Fuminobu",
    title = "{Moduli-induced gravitino problem}",
    eprint = "hep-ph/0602061",
    archivePrefix = "arXiv",
    reportNumber = "DESY-06-014",
    doi = "10.1103/PhysRevLett.96.211301",
    journal = "Phys. Rev. Lett.",
    volume = "96",
    pages = "211301",
    year = "2006"
}

@article{Shuhmaher:2007pv,
    author = "Shuhmaher, Natalia",
    title = "{A Note on the moduli-induced gravitino problem}",
    eprint = "hep-ph/0703319",
    archivePrefix = "arXiv",
    doi = "10.1088/1126-6708/2008/12/094",
    journal = "JHEP",
    volume = "12",
    pages = "094",
    year = "2008"
}

@article{Akita:2016usy,
    author = "Akita, Kensuke and Kobayashi, Tatsuo and Oikawa, Akane and Otsuka, Hajime",
    title = "{Moduli mediation without moduli-induced gravitino problem}",
    eprint = "1603.08399",
    archivePrefix = "arXiv",
    primaryClass = "hep-ph",
    reportNumber = "WU-HEP-16-06, EPHOU-16-001",
    doi = "10.1007/JHEP05(2016)178",
    journal = "JHEP",
    volume = "05",
    pages = "178",
    year = "2016"
}

@article{Dutta:2009uf,
    author = "Dutta, Bhaskar and Leblond, Louis and Sinha, Kuver",
    title = "{Mirage in the Sky: Non-thermal Dark Matter, Gravitino Problem, and Cosmic Ray Anomalies}",
    eprint = "0904.3773",
    archivePrefix = "arXiv",
    primaryClass = "hep-ph",
    reportNumber = "MIFP-09-19, NSF-KITP-09-52",
    doi = "10.1103/PhysRevD.80.035014",
    journal = "Phys. Rev. D",
    volume = "80",
    pages = "035014",
    year = "2009"
}

@article{Kawasaki:2006hm,
    author = "Kawasaki, Masahiro and Takahashi, Fuminobu and Yanagida, T. T.",
    title = "{The Gravitino-overproduction problem in inflationary universe}",
    eprint = "hep-ph/0605297",
    archivePrefix = "arXiv",
    reportNumber = "DESY-06-072",
    doi = "10.1103/PhysRevD.74.043519",
    journal = "Phys. Rev. D",
    volume = "74",
    pages = "043519",
    year = "2006"
}

@article{Kawasaki:1994af,
    author = "Kawasaki, M. and Moroi, T.",
    title = "{Gravitino production in the inflationary universe and the effects on big bang nucleosynthesis}",
    eprint = "hep-ph/9403364",
    archivePrefix = "arXiv",
    reportNumber = "ICRR-315-94-10, TU-457",
    doi = "10.1143/PTP.93.879",
    journal = "Prog. Theor. Phys.",
    volume = "93",
    pages = "879--900",
    year = "1995"
}

@mastersthesis{Moroi:1995fs,
    author = "Moroi, Takeo",
    title = "{Effects of the gravitino on the inflationary universe}",
    eprint = "hep-ph/9503210",
    archivePrefix = "arXiv",
    reportNumber = "TU-479",
    type = "Other thesis",
    month = "3",
    year = "1995"
}

@article{Kohri:2005wn,
    author = "Kohri, Kazunori and Moroi, Takeo and Yotsuyanagi, Akira",
    title = "{Big-bang nucleosynthesis with unstable gravitino and upper bound on the reheating temperature}",
    eprint = "hep-ph/0507245",
    archivePrefix = "arXiv",
    reportNumber = "OU-TAP-261, TU-749",
    doi = "10.1103/PhysRevD.73.123511",
    journal = "Phys. Rev. D",
    volume = "73",
    pages = "123511",
    year = "2006"
}

@article{Kawasaki:2008qe,
    author = "Kawasaki, Masahiro and Kohri, Kazunori and Moroi, Takeo and Yotsuyanagi, Akira",
    title = "{Big-Bang Nucleosynthesis and Gravitino}",
    eprint = "0804.3745",
    archivePrefix = "arXiv",
    primaryClass = "hep-ph",
    reportNumber = "TU-812",
    doi = "10.1103/PhysRevD.78.065011",
    journal = "Phys. Rev. D",
    volume = "78",
    pages = "065011",
    year = "2008"
}

@article{Pradler:2006qh,
    author = "Pradler, Josef and Steffen, Frank Daniel",
    title = "{Thermal gravitino production and collider tests of leptogenesis}",
    eprint = "hep-ph/0608344",
    archivePrefix = "arXiv",
    reportNumber = "MPP-2006-101",
    doi = "10.1103/PhysRevD.75.023509",
    journal = "Phys. Rev. D",
    volume = "75",
    pages = "023509",
    year = "2007"
}

@article{Blinov:2014nla,
    author = "Blinov, Nikita and Kozaczuk, Jonathan and Menon, Arjun and Morrissey, David E.",
    title = "{Confronting the moduli-induced lightest-superpartner problem}",
    eprint = "1409.1222",
    archivePrefix = "arXiv",
    primaryClass = "hep-ph",
    doi = "10.1103/PhysRevD.91.035026",
    journal = "Phys. Rev. D",
    volume = "91",
    number = "3",
    pages = "035026",
    year = "2015"
}

@article{Baer:2021zbj,
    author = "Baer, Howard and Barger, Vernon and Deal, Robert Wiley",
    title = "{An anthropic solution to the cosmological moduli problem}",
    eprint = "2111.05971",
    archivePrefix = "arXiv",
    primaryClass = "hep-ph",
    reportNumber = "OU-HEP-211110",
    doi = "10.1016/j.jheap.2022.03.005",
    journal = "JHEAp",
    volume = "34",
    pages = "33--39",
    year = "2022"
}

@article{Bae:2022okh,
    author = "Bae, Kyu Jung and Baer, Howard and Barger, Vernon and Deal, Robert Wiley",
    title = "{The cosmological moduli problem and naturalness}",
    eprint = "2201.06633",
    archivePrefix = "arXiv",
    primaryClass = "hep-ph",
    reportNumber = "OU-HEP-211030",
    doi = "10.1007/JHEP02(2022)138",
    journal = "JHEP",
    volume = "02",
    pages = "138",
    year = "2022"
}

@article{Baer:2023bbn,
    author = "Baer, Howard and Barger, Vernon and Wiley Deal, Robert",
    title = "{Dark matter and dark radiation from the early universe with a modulus coupled to the PQMSSM}",
    eprint = "2301.12546",
    archivePrefix = "arXiv",
    primaryClass = "hep-ph",
    reportNumber = "OU-HEP-230204",
    doi = "10.1007/JHEP06(2023)083",
    journal = "JHEP",
    volume = "06",
    pages = "083",
    year = "2023"
}

@phdthesis{WileyDeal:2023sry,
    author = "Wiley Deal, Robert",
    title = "{The dark universe: the interplay of cosmological moduli, axions, and the MSSM}",
    school = "Oklahoma U.",
    month = "5",
    year = "2023"
}

@article{Paige:2003mg,
    author = "Paige, Frank E. and Protopopescu, Serban D. and Baer, Howard and Tata, Xerxes",
    title = "{ISAJET 7.69: A Monte Carlo event generator for pp, anti-p p, and e+e- reactions}",
    eprint = "hep-ph/0312045",
    archivePrefix = "arXiv",
    month = "12",
    year = "2003"
}

@article{ATLAS:2024lda,
    author = "Aad, Georges and others",
    collaboration = "ATLAS",
    title = "{The quest to discover supersymmetry at the ATLAS experiment}",
    eprint = "2403.02455",
    archivePrefix = "arXiv",
    primaryClass = "hep-ex",
    reportNumber = "CERN-EP-2024-056",
    doi = "10.1016/j.physrep.2024.09.010",
    journal = "Phys. Rept.",
    volume = "1116",
    pages = "261--300",
    year = "2025"
}

@article{Sekmen:2025bxv,
    author = "Sekmen, Sezen",
    title = "{Supersymmetry Searches in CMS Run 2: A Complete Review}",
    eprint = "2510.17971",
    archivePrefix = "arXiv",
    primaryClass = "hep-ex",
    doi = "10.53941/hihep.2025.100018",
    journal = "HiHEP",
    volume = "1",
    number = "2",
    pages = "18",
    year = "2025"
}

@article{Bae:2013hma,
    author = "Bae, Kyu Jung and Baer, Howard and Chun, Eung Jin",
    title = "{Mixed axion/neutralino dark matter in the SUSY DFSZ axion model}",
    eprint = "1309.5365",
    archivePrefix = "arXiv",
    primaryClass = "hep-ph",
    doi = "10.1088/1475-7516/2013/12/028",
    journal = "JCAP",
    volume = "12",
    pages = "028",
    year = "2013"
}

@article{Baer:2002fv,
    author = "Baer, Howard and Balazs, Csaba and Belyaev, Alexander",
    title = "{Neutralino relic density in minimal supergravity with coannihilations}",
    eprint = "hep-ph/0202076",
    archivePrefix = "arXiv",
    reportNumber = "FSU-HEP-020208",
    doi = "10.1088/1126-6708/2002/03/042",
    journal = "JHEP",
    volume = "03",
    pages = "042",
    year = "2002"
}

@article{Moroi:1999zb,
    author = "Moroi, Takeo and Randall, Lisa",
    title = "{Wino cold dark matter from anomaly mediated SUSY breaking}",
    eprint = "hep-ph/9906527",
    archivePrefix = "arXiv",
    reportNumber = "IASSNS-HEP-99-54, PUPT-1873, MIT-CTP-2873",
    doi = "10.1016/S0550-3213(99)00748-8",
    journal = "Nucl. Phys. B",
    volume = "570",
    pages = "455--472",
    year = "2000"
}

@article{Randall:1998uk,
    author = "Randall, Lisa and Sundrum, Raman",
    title = "{Out of this world supersymmetry breaking}",
    eprint = "hep-th/9810155",
    archivePrefix = "arXiv",
    reportNumber = "MIT-CTP-2788, PUPT-1815, BUHEP-98-26",
    doi = "10.1016/S0550-3213(99)00359-4",
    journal = "Nucl. Phys. B",
    volume = "557",
    pages = "79--118",
    year = "1999"
}

@article{Giudice:1998xp,
    author = "Giudice, Gian F. and Luty, Markus A. and Murayama, Hitoshi and Rattazzi, Riccardo",
    title = "{Gaugino mass without singlets}",
    eprint = "hep-ph/9810442",
    archivePrefix = "arXiv",
    reportNumber = "CERN-TH-98-337, LBNL-42419, LBL-42419, UCB-PTH-98-50, UMD-PP-99-037",
    doi = "10.1088/1126-6708/1998/12/027",
    journal = "JHEP",
    volume = "12",
    pages = "027",
    year = "1998"
}

@article{Baer:2022fou,
    author = "Baer, Howard and Barger, Vernon and Deal, Robert Wiley",
    title = "{On dark radiation from string moduli decay to ALPs}",
    eprint = "2204.01130",
    archivePrefix = "arXiv",
    primaryClass = "hep-ph",
    reportNumber = "OU-HEP-220401",
    doi = "10.1016/j.jheap.2022.04.001",
    journal = "JHEAp",
    volume = "34",
    pages = "40--48",
    year = "2022"
}

@article{Kim:1983dt,
    author = "Kim, Jihn E. and Nilles, Hans Peter",
    title = "{The mu Problem and the Strong CP Problem}",
    reportNumber = "UGVA-DPT 1983/10-410",
    doi = "10.1016/0370-2693(84)91890-2",
    journal = "Phys. Lett. B",
    volume = "138",
    pages = "150--154",
    year = "1984"
}

@article{Bae:2019dgg,
    author = "Bae, Kyu Jung and Baer, Howard and Barger, Vernon and Sengupta, Dibyashree",
    title = "{Revisiting the SUSY $\mu$ problem and its solutions in the LHC era}",
    eprint = "1902.10748",
    archivePrefix = "arXiv",
    primaryClass = "hep-ph",
    reportNumber = "CTPU-PTC-19-06",
    doi = "10.1103/PhysRevD.99.115027",
    journal = "Phys. Rev. D",
    volume = "99",
    number = "11",
    pages = "115027",
    year = "2019"
}

@article{Krauss:1988zc,
    author = "Krauss, Lawrence M. and Wilczek, Frank",
    title = "{Discrete Gauge Symmetry in Continuum Theories}",
    reportNumber = "YCTP-P26-88, NSF-ITP-88-187",
    doi = "10.1103/PhysRevLett.62.1221",
    journal = "Phys. Rev. Lett.",
    volume = "62",
    pages = "1221",
    year = "1989"
}

@article{Ibanez:1991pr,
    author = "Ibanez, Luis E. and Ross, Graham G.",
    title = "{Discrete gauge symmetries and the origin of baryon and lepton number conservation in supersymmetric versions of the standard model}",
    reportNumber = "CERN-TH-6111-91",
    doi = "10.1016/0550-3213(92)90195-H",
    journal = "Nucl. Phys. B",
    volume = "368",
    pages = "3--37",
    year = "1992"
}

@article{Dreiner:2005rd,
    author = "Dreiner, Herbi K. and Luhn, Christoph and Thormeier, Marc",
    title = "{What is the discrete gauge symmetry of the MSSM?}",
    eprint = "hep-ph/0512163",
    archivePrefix = "arXiv",
    doi = "10.1103/PhysRevD.73.075007",
    journal = "Phys. Rev. D",
    volume = "73",
    pages = "075007",
    year = "2006"
}

@article{Lee:2011dya,
    author = "Lee, Hyun Min and Raby, Stuart and Ratz, Michael and Ross, Graham G. and Schieren, Roland and Schmidt-Hoberg, Kai and Vaudrevange, Patrick K. S.",
    title = "{Discrete R symmetries for the MSSM and its singlet extensions}",
    eprint = "1102.3595",
    archivePrefix = "arXiv",
    primaryClass = "hep-ph",
    reportNumber = "TUM-HEP-793-11, LMU-ASC-06-11, OHSTPY-HEP-T-11-001, CERN-PH-TH-2011-022, OUTP-11-33P",
    doi = "10.1016/j.nuclphysb.2011.04.009",
    journal = "Nucl. Phys. B",
    volume = "850",
    pages = "1--30",
    year = "2011"
}

@article{Baer:2018avn,
    author = "Baer, Howard and Barger, Vernon and Sengupta, Dibyashree",
    title = "{Gravity safe, electroweak natural axionic solution to strong $CP$ and SUSY $\mu$ problems}",
    eprint = "1810.03713",
    archivePrefix = "arXiv",
    primaryClass = "hep-ph",
    reportNumber = "OU-HEP-180930",
    doi = "10.1016/j.physletb.2019.01.007",
    journal = "Phys. Lett. B",
    volume = "790",
    pages = "58--63",
    year = "2019"
}

@article{Bhattiprolu:2021rrj,
    author = "Bhattiprolu, Prudhvi N. and Martin, Stephen P.",
    title = "{High-quality axions in solutions to the {\ensuremath{\mu}} problem}",
    eprint = "2106.14964",
    archivePrefix = "arXiv",
    primaryClass = "hep-ph",
    doi = "10.1103/PhysRevD.104.055014",
    journal = "Phys. Rev. D",
    volume = "104",
    number = "5",
    pages = "055014",
    year = "2021"
}

@article{Baer:2025oid,
    author = "Baer, Howard and Barger, Vernon and Sengupta, Dibyashree and Zhang, Kairui",
    title = "{All axion dark matter from supersymmetric models}",
    eprint = "2502.06955",
    archivePrefix = "arXiv",
    primaryClass = "hep-ph",
    reportNumber = "OU-HEP-250204",
    doi = "10.1103/1n3m-g69m",
    journal = "Phys. Rev. D",
    volume = "111",
    number = "11",
    pages = "L111702",
    year = "2025"
}

@article{Baer:2025srs,
    author = "Baer, Howard and Barger, Vernon and Bolich, Jessica and Sengupta, Dibyashree and Zhang, Kairui",
    title = "{Aspects of the WIMP quality problem and R-parity violation in natural supersymmetry with all axion dark~matter}",
    eprint = "2505.09785",
    archivePrefix = "arXiv",
    primaryClass = "hep-ph",
    reportNumber = "OU-HEP-250509",
    doi = "10.1088/1475-7516/2025/10/072",
    journal = "JCAP",
    volume = "10",
    pages = "072",
    year = "2025"
}

@article{Nilles:2017heg,
    author = "Nilles, Hans Peter",
    title = "{Stringy Origin of Discrete R-symmetries}",
    eprint = "1705.01798",
    archivePrefix = "arXiv",
    primaryClass = "hep-ph",
    doi = "10.22323/1.292.0017",
    journal = "PoS",
    volume = "CORFU2016",
    pages = "017",
    year = "2017"
}

@article{Chen:2012tia,
    author = "Chen, Mu-Chun and Fallbacher, Maximilian and Ratz, Michael",
    editor = "Szczerbinska, Barbara and Babu, Kaladi and Balantekin, Baha and Dutta, Bhaskar and Mohapatra, Rabindra N.",
    title = "{Supersymmetric unification and R symmetries}",
    eprint = "1211.6247",
    archivePrefix = "arXiv",
    primaryClass = "hep-ph",
    reportNumber = "UCI-TR-2012-19, TUM-HEP-869-12, FLAVOR-EU-32, CETUP-016",
    doi = "10.1142/S0217732312300443",
    journal = "Mod. Phys. Lett. A",
    volume = "27",
    pages = "1230044",
    year = "2012"
}

@article{Balasubramanian:2005zx,
    author = "Balasubramanian, Vijay and Berglund, Per and Conlon, Joseph P. and Quevedo, Fernando",
    title = "{Systematics of moduli stabilisation in Calabi-Yau flux compactifications}",
    eprint = "hep-th/0502058",
    archivePrefix = "arXiv",
    reportNumber = "DAMTP-2005-10, UNH-05-01, UPR-1109-T",
    doi = "10.1088/1126-6708/2005/03/007",
    journal = "JHEP",
    volume = "03",
    pages = "007",
    year = "2005"
}

@article{LZ:2024zvo,
    author = "Aalbers, J. and others",
    collaboration = "LZ",
    title = "{Dark Matter Search Results from 4.2{\,}{\,}Tonne-Years of Exposure of the LUX-ZEPLIN (LZ) Experiment}",
    eprint = "2410.17036",
    archivePrefix = "arXiv",
    primaryClass = "hep-ex",
    reportNumber = "FERMILAB-PUB-24-0796-V",
    doi = "10.1103/4dyc-z8zf",
    journal = "Phys. Rev. Lett.",
    volume = "135",
    number = "1",
    pages = "011802",
    year = "2025"
}

@article{Ellis:2002wv,
    author = "Ellis, John R. and Olive, Keith A. and Santoso, Yudi",
    title = "{The MSSM parameter space with nonuniversal Higgs masses}",
    eprint = "hep-ph/0204192",
    archivePrefix = "arXiv",
    reportNumber = "CERN-TH-2002-081, UMN-TH-2049-02, TPI-MINN-02-09",
    doi = "10.1016/S0370-2693(02)02071-3",
    journal = "Phys. Lett. B",
    volume = "539",
    pages = "107--118",
    year = "2002"
}

@article{Baer:2005bu,
    author = "Baer, Howard and Mustafayev, Azar and Profumo, Stefano and Belyaev, Alexander and Tata, Xerxes",
    title = "{Direct, indirect and collider detection of neutralino dark matter in SUSY models with non-universal Higgs masses}",
    eprint = "hep-ph/0504001",
    archivePrefix = "arXiv",
    reportNumber = "FSU-HEP-050315, UH-511-1067-05",
    doi = "10.1088/1126-6708/2005/07/065",
    journal = "JHEP",
    volume = "07",
    pages = "065",
    year = "2005"
}

@article{Baer:2026wre,
    author = "Baer, Howard and Barger, Vernon and Zhang, Kairui",
    title = "{Natural SUSY with mixed axion/axino dark matter}",
    eprint = "2604.04687",
    archivePrefix = "arXiv",
    primaryClass = "hep-ph",
    reportNumber = "OU-HEP-260401",
    doi = "10.1016/j.physletb.2026.140655",
    journal = "Phys. Lett. B",
    volume = "879",
    pages = "140655",
    year = "2026"
}

@article{WileyDeal:2025wgh,
    author = "Wiley Deal, Robert and Barrowes, Leia and Giblin, John T. and Sinha, Kuver and Watson, Scott and Adams, Fred C.",
    title = "{Cosmological moduli and non-perturbative production of axions}",
    eprint = "2501.17229",
    archivePrefix = "arXiv",
    primaryClass = "hep-ph",
    doi = "10.1007/JHEP07(2025)154",
    journal = "JHEP",
    volume = "07",
    pages = "154",
    year = "2025"
}

@article{Price:2026kjs,
    author = "Price, Leia and Sinha, Kuver and Wiley Deal, Robert",
    title = "{Global Asymptotics, the Swampland Conjectures, and Preheating of String Moduli}",
    eprint = "2607.18442",
    archivePrefix = "arXiv",
    primaryClass = "hep-th",
    month = "7",
    year = "2026"
}

@article{Antusch:2017flz,
    author = "Antusch, Stefan and Cefala, Francesco and Krippendorf, Sven and Muia, Francesco and Orani, Stefano and Quevedo, Fernando",
    title = "{Oscillons from String Moduli}",
    eprint = "1708.08922",
    archivePrefix = "arXiv",
    primaryClass = "hep-th",
    doi = "10.1007/JHEP01(2018)083",
    journal = "JHEP",
    volume = "01",
    pages = "083",
    year = "2018"
}

@article{Leedom:2024qgr,
    author = "Leedom, Jacob M. and Putti, Margherita and Righi, Nicole and Westphal, Alexander",
    title = "{Preheating axions in string cosmology}",
    eprint = "2411.18496",
    archivePrefix = "arXiv",
    primaryClass = "hep-th",
    reportNumber = "DESY-24-163, KCL-PH-TH/2024-60",
    doi = "10.1007/JHEP04(2025)095",
    journal = "JHEP",
    volume = "04",
    pages = "095",
    year = "2025"
}

@article{Giblin:2017wlo,
    author = "Giblin, John T. and Kane, Gordon and Nesbit, Eva and Watson, Scott and Zhao, Yue",
    title = "{Was the Universe Actually Radiation Dominated Prior to Nucleosynthesis?}",
    eprint = "1706.08536",
    archivePrefix = "arXiv",
    primaryClass = "hep-th",
    doi = "10.1103/PhysRevD.96.043525",
    journal = "Phys. Rev. D",
    volume = "96",
    number = "4",
    pages = "043525",
    year = "2017"
}
\bibliographystyle{elsarticle-num}

\end{document}